\documentclass[smallextended]{svjour3}       
\usepackage{lineno,hyperref}
\usepackage{graphicx}
\usepackage{amsmath}
\usepackage{amssymb}
\usepackage{amsfonts}
\usepackage{array}
\usepackage[caption=false,font=footnotesize]{subfig}
\usepackage{fixltx2e}
\usepackage{url}
\usepackage{tikz}
\usetikzlibrary{calc}
\usepackage{tkz-graph}
 \makeatletter
 \renewcommand*{\Edge}[1][]{\tkz@edge[#1]}%
 \def\tkz@edge[#1](#2)(#3){%
 \setkeys[GR]{edge}{#1}%
  \begingroup%
 \ifthenelse{\equal{\cmdGR@edge@double}{}}{%
 \tikzset{LocalEdgeStyle/.style={color = \cmdGR@edge@color,
                                 line width = \cmdGR@edge@lw}}}{%
 \tikzset{LocalEdgeStyle/.style={line width = \cmdGR@edge@dd,
                                 color = \cmdGR@edge@double,
                                 double distance = \cmdGR@edge@lw,
                                 double  = \cmdGR@edge@color}}}%
 \ifGR@edge@local%
       \tikzset{EdgeStyle/.style={}}%
       \fi
    \ifthenelse{\equal{\cmdGR@edge@label}{}}{%
      \protected@edef\@tempa{%
      \noexpand   \draw[LocalEdgeStyle,\cmdGR@edge@style,EdgeStyle]}%
                  \@tempa (#2) to (#3)}{%
      \protected@edef\@tempa{%
      \noexpand   \draw[LocalEdgeStyle,\cmdGR@edge@style,EdgeStyle] (#2) to%
     node[fill = \cmdGR@edge@labelcolor,
          text = \cmdGR@edge@labeltext,
          \cmdGR@edge@labelstyle,LabelStyle]}\@tempa
    {\cmdGR@edge@label} (#3)}%
    ;
 \endgroup%
 }%
 \makeatother
 \usepackage{tikz-qtree}
\usepackage{mathtools}
\usepackage{algorithm}
\usepackage{algorithmic}
\usepackage{enumitem}
\usepackage{subfig}
\usepackage{adjustbox}
\usepackage{booktabs}
\usepackage{cite}

\tikzstyle{snofill}=[circle,draw,font={A}]
\tikzstyle{bfill}=[circle,draw,preaction={fill=black!20}]
\tikzstyle{rfill}=[circle,draw,preaction={fill=red!20}]
\tikzstyle{rrfill}=[circle,draw,preaction={fill=red!40}]
\tikzstyle{rrrfill}=[rectangle ,draw,preaction={fill=red!20}]

\usetikzlibrary{mindmap,trees}

\newcommand{\etal}{\mbox{\emph{et al.\ }}}

\newcommand{\bigo}[1]{{\cal O} (#1)}

\begin{document}

\title{MapReduce Particle Filtering with Exact Resampling and Deterministic Runtime}

\author{
 Jeyarajan Thiyagalingam \and
 Lykourgos Kekempanos \and
 Simon Maskell
}

\institute{
  Department of Electrical Engineering and Electronics, \\
  University of Liverpool,
  Liverpool, L69 3GJ, UK.\\
  \email{\{T.Jeyarajan, L.Kekempanos, S.Maskell\}@liverpool.ac.uk}
}
\maketitle

\begin{abstract}
Particle filtering is a numerical Bayesian technique that has great
potential for solving sequential estimation problems involving
non-linear and non-Gaussian models. Since the estimation accuracy
achieved by particle filters improves as the number of particles
increases, it is natural to consider as many particles as
possible. MapReduce is a generic programming model that makes it
possible to scale a wide variety of algorithms to Big data. However,
despite the application of particle filters across many domains,
little attention has been devoted to implementing particle filters
using MapReduce.

In this paper, we describe an implementation of a particle filter
using MapReduce. We focus on a component that what would otherwise be
a bottleneck to parallel execution, the resampling component. We
devise a new implementation of this component, which requires no
approximations, has $O\left(N\right)$ spatial complexity and
deterministic $O\left(\left(\log N\right)^2\right)$ time
complexity. Results demonstrate the utility of this new component and
culminate in consideration of a particle filter with $2^{24}$
particles being distributed across $512$ processor cores.
\end{abstract}



\section{Introduction}
\label{sec:intro}

Particle filters are a Bayesian Monte-Carlo method that provide a
general framework for estimation in response to an incoming stream of
data.  The key idea is to represent the probability density function
(pdf) of the state of a system using random samples (known as
particles). These samples are propagated across iterations in time in
a way that capitalises on an application-specific non-linear,
non-Gaussian state-space model. This state-space model describes both
the dynamic evolution of the state and the relationship between the
state and the measurements.  The use of random samples to articulate
uncertainty means that particle filters can be applied to a variety of
real-world problems without any need to approximate the models
used. This is in contrast to alternative techniques (e.g., the Extended
Kalman filter, EKF) that approximate the models such that the
uncertainty present can be approximated using a parametric probability
density (a multivariate Gaussian in the case of an EKF). The result is
that a particle filter typically outperforms such alternative
techniques in scenarios involving pronounced departures from
linear-Gaussian models. Such scenarios are widespread. This is
arguably the reason why particle filters, since their
inception~\cite{Gordon:1993}, have been applied successfully in such a
diverse range of
contexts~\cite{Thrun:2001,Sakaki:2010,Creal:2012,Gustafsson:2002}.

Particle filters have the appealing property that, as the number of
samples increases, the ability of the samples to represent the pdf
increases and the accuracy of estimates derived from the particles
improves: an upper-bound on the variance of an estimate scales as
$O\left(\frac{1}{N}\right)$. It is therefore natural to seek to use as
many particles as possible. However, when the number of samples
becomes very large, the samples will not physically fit within the
memory space of a single compute node. Big data platforms have been
developed to address the generic problem of which this is a special
case. These platforms work by identifying abstractions of algorithms
that make the potential for parallelism apparent. The platforms (and
not the programmer) are then able to exploit the available
computational resources to distribute the processing. One popular
abstraction is MapReduce (which is described in more detail in
section~\ref{sec:mapreduce}). Various techniques have been
developed to distribute particle filters across multiple
processor-cores (see section~\ref{sec:related} for the details),
but MapReduce has not been used with particle filters extensively
(that said, \cite{Bai:2016} and~\cite{Bai:2013} are counter-examples we are aware of).

The resampling component is a critical component of a particle filter
and non-trivial to parallelise. As will be discussed in more detail in
section~\ref{sec:related}, previous approaches to distributing the
resampling step have focused on modifying the resampling process with
the aim of making it more amenable to distributed implementation. One
notable exception exists\cite{Maskell:SIMDPF2006}\footnote{Though we are not aware of any empirical analysis of this
approach being published.} and ensures that the
output from the distributed implementation is exactly that output from
a single-processor implementation while also ensuring deterministic
data transfer and runtime. Such deterministic runtime is important in real-time applications (which are widespread) where the
output of the particle filter is used to feed the input of another process, which needs to receive that input within a specified latency.

In this paper, we present an improved parallel implementation strategy
for the resampling component, a MapReduce representation of the
particle filter (including this resampling component) and instantiate the
particle filter in the context of two Big data platforms. In doing so,
this paper makes the following key contributions:
\begin{itemize}
\item We propose an improved implementation of an exact deterministic
  resampling algorithm that has better temporal complexity compared to
  the current state-of-the-art\cite{Maskell:SIMDPF2006}. More
  specifically, the proposed version of the parallel algorithm has the
  complexity of ${\cal O}((\log_2N)^2)$ compared to the original
  complexity of ${\cal O}((\log_2N)^3)$.
\item We provide two different MapReduce variants of our new algorithm
  that fit both with the in-memory processing and out-of-core
  processing models. These are the processing models used by Hadoop
  and Spark respectively.
\item We perform detailed performance and scalability analysis of our
  new algorithm in comparison to both the pre-existing
  state-of-the-art\cite{Maskell:SIMDPF2006} and an implementation
  optimised for a single processor-core. We deliberately chose an application that stresses the resampling component of the particle filter such that our analysis relates to worst-case performance.
\end{itemize}

The remainder of this paper is organised as follows: In
section~\ref{sec:bigdata}, we provide a brief overview of Big data
processing, and the MapReduce programming model. This is then followed
by a detailed description of particle filtering in
section~\ref{sec:particlefiltering}. In section~\ref{sec:components},
we describe the fundamental building blocks that are used to construct
the implementations of the particle filtering algorithm, including, in section~\ref{sec:improvement},
the new component of the resampling algorithm.  We then describe our MapReduce-based
particle filtering implementation in section~\ref{sec:mrpf}. We follow
this section with an evaluation of our algorithms on key two MapReduce
frameworks in Section~\ref{sec:evaluation}. Section~\ref{sec:related}
highlights related work before section~\ref{sec:conclusions}
concludes.

\section{Big data Processing}
\label{sec:bigdata}

The focus in this paper is on the problem of using large numbers of
samples within a particle filter. Big data processing frameworks (e.g.,
Apache projects such as: Hadoop~\cite{Hadoop:Main},
Spark~\cite{Spark:Main} and Storm~\cite{Storm:Main}\footnote{Including
  the associated ever-growing ecosystem of tools (e.g.,
  Mahout~\cite{Mahout:Main} and GraphX for Spark~\cite{Singh:2014}).})
are designed for handling large amounts of data and can therefore be
applied in this context\footnote{Conventional High Performance
  Computing (HPC) approaches use parallel computations to optimise
  processing time. We refer the reader
  to~\cite{Asanovic:2009:Landscape} for a good coverage of HPC-bound
  approaches for parallelising applications.}. We therefore focus in
this paper on using such frameworks in conjunction with parallel
computational resources, such as clusters, to handle large volumes of
data\footnote{We anticipate that the `heat wall' (i.e., the inability to remove enough heat from transistors that switch ever faster) will mean that for chip manufacturers to meet the expectation set by Moore's law, they will soon (if not already) be doubling the number of cores (not transistors per square inch) used in each processor each year. In ten years' time, if this trend continues, we would have desktop computers with a thousand times as many cores as today. This trend motivates the authors to design implementation strategies for particle filters that are well suited to the multi-core processors which will, we believe, become increasingly prevalent over time.}. In this section, we discuss the use of such Big data frameworks
in general and, in particular, one of the programming models that
underpins such frameworks, the MapReduce programming model.

\subsection{Big data Frameworks}
\label{sec:subsec:bigdataframeworks}

An attractive approach for scaling the problem with data is to use Big
data frameworks. More strictly, Big data frameworks go beyond the
issue of data volume and address much wider issues covering augmented
V's of data, for instance {\em volume, velocity, variety, value} and
{\em veracity}~\cite{Schroeck:2012}.  Big data framework-based
solutions are process-centric: the programmer describes the algorithm
in a way that enables the framework itself to understand (and attempt
to exploit) the potential to distribute the data and
processing\footnote{This contrasts HPC-based solutions, where the
  programmer aims to exploit intricate knowledge of the underlying
  architecture to ensure that data movement and processing are jointly
  optimised for the specific hardware.}. The result of this delegation
of the optimisation for speed to the framework is that, while many of
today's Big data frameworks can handle large volumes of data, none can
match the runtime performance of conventional HPC
systems~\cite{2015:HPCHadoop}.

There are a growing number of different programming models that are
used to describe algorithms within Big data frameworks. These models
include MapReduce~\cite{MapReduce:Dean:2008}, Stream
Processing~\cite{Herath:StreamFlow,Spark:Main,Storm:Main} and
Query-based techniques~\cite{Wang:2015,PigLatin:Main}. Here, we focus
on one such programming model, MapReduce.

\subsection{The MapReduce Programming Model}
\label{sec:mapreduce}

\begin{figure}
  \centering
  \includegraphics[width=0.7\textwidth]{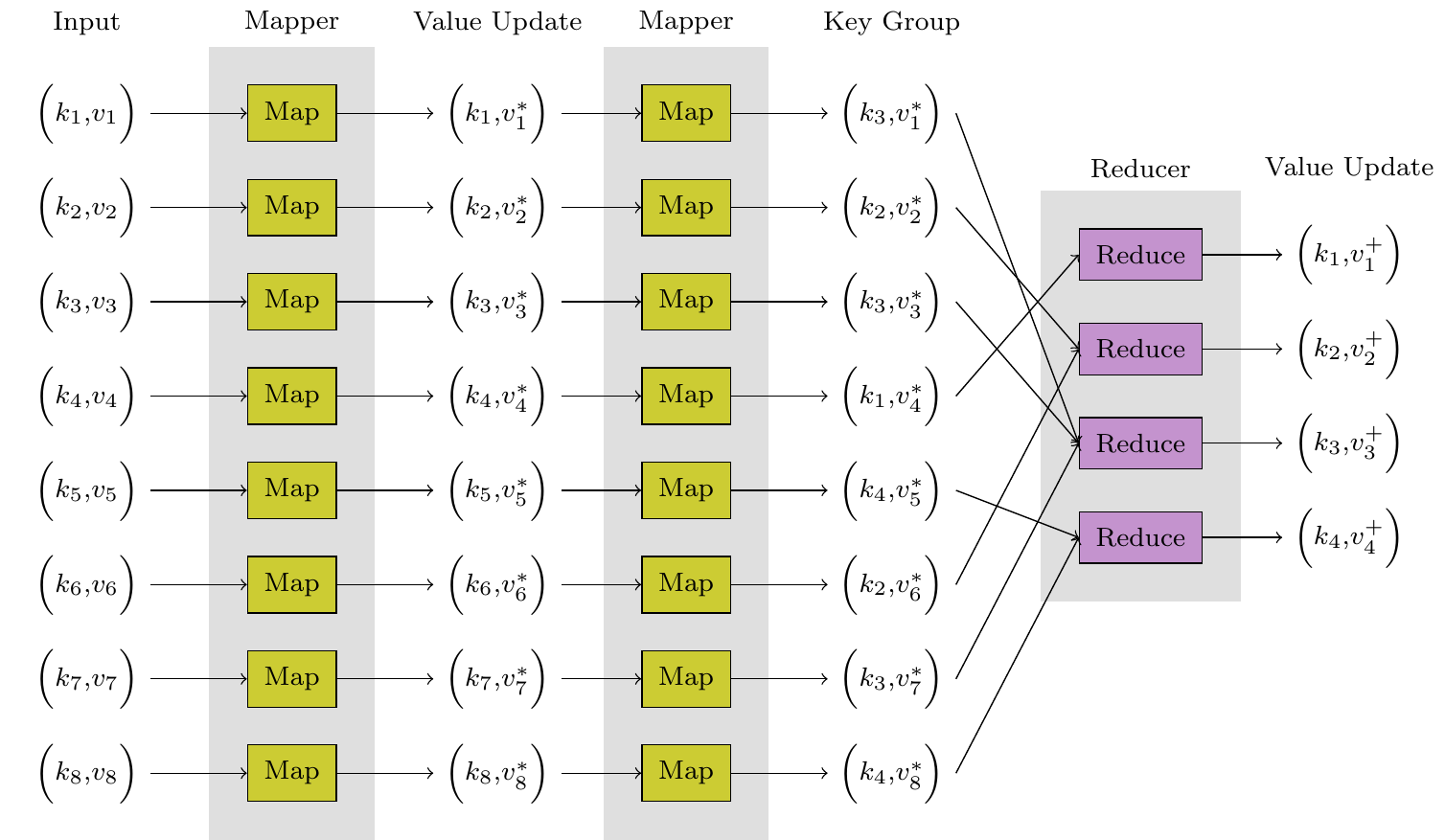}
  \caption{General MapReduce Processing Model.}
  \label{fig:mapreduce}
\end{figure}

MapReduce is a popular programming model used in many Big data
processing frameworks (and even some HPC frameworks). The key focus of
the MapReduce model is on enabling the framework to distribute the
processing of a large dataset by expressing algorithms in terms of
{\em map} and {\em reduce} operations, via defining {\em mappers} and
{\em reducers}. Mappers, when applied to each datum, output a list of
{\em (key, value)} pairs. The framework then collates all the values
associated with each key. Reducers are then applied to the list of
values for each key to output a single value. Note that both the map
and reduce operations are inherently parallel across all data and keys
respectively\footnote{The exact number of mapper and reducer processes
  on a parallel resource (for instance, a multi-node cluster) varies
  depending on the configuration, but the important point is that the
  algorithm developer does not need to worry about how the processes
  are distributed when defining the algorithm. Of course, that does not
  mean that there is not utility in the developer describing algorithms
  using mappers and reducers that are well suited to the problem being
  tackled and to the configuration being used.}. To exemplify this,
consider a dataset where each datum is a sentence in a Big document
(e.g., Wikipedia). The problem of counting the total number of
occurrences of each word in the document corpus can then be described
as using the words as the key, a mapper that outputs a (non-zero)
count of the number of times each word occurs in each
sentence\footnote{Note that the output from each sentence would only
  be for the words that occur in that sentence, not every word that
  ever occurs in the corpus.} and a reducer that calculates the sum of
the counts. For each word, the reducer's output is then the sum over
all sentences of the counts per sentence. Another example is shown in
Figure~\ref{fig:mapreduce} and illustrates the ability to pass
(key,~value) pairs into a mapper and thereby use the output of one
mapper as the input into a second mapper.

Two key frameworks that support MapReduce, albeit in slightly
different ways, are Hadoop and Spark. These are now considered in
turn.

\subsubsection{Hadoop}

MapReduce is one of the two fundamental components of Hadoop. The
other is the Hadoop Distributed File System (HDFS). HDFS enables
multiple computers' disks to be accessed in much the same way as if it
were a single (Big) disk. In Hadoop, the mapper and reducer generate
files which are stored in HDFS, such that Hadoop implements data
movement entirely via the file system.

\subsubsection{Spark}

The Spark framework operates using a different principle. First, at
the Application Programming Interface (API) level, Spark provides a distributed data structure known as a
Resilient Distributed Dataset (RDDs)~\cite{SparkRDD:2012}. MapReduce
is then just one of a large number of {\em transformations} that (via
a rich set of APIs) can be applied to RDDs. It is also important to
realise that evaluations in Spark are {\em lazily} executed. This
means, unlike conventional processing engines (e.g., Hadoop), executions
never actually happen when transformations are defined. Instead,
transformations are used to compose a data-flow graph and execution
happens when forced through {\em actions} (i.e., when necessary). This
delayed evaluation enables the Spark framework to optimise (and plan)
the execution\footnote{This can make it hard for a programmer to debug
  algorithmic implementations, particularly if the programmer is
  unfamiliar with debugging software performing lazy evaluation.}. The
result is often significant improvements in runtime
performance. Another important property of RDDs is that they can
reside in memory, disk or in combination.  Indeed, although Spark can
make use of HDFS, the data movements in Spark are primarily via
memory. Again, this can result in significant improvements in runtime
performance relative to Hadoop.
\section{Particle Filtering}
\label{sec:particlefiltering}

We now provide a brief description of particle filtering. The reader unfamiliar with particle filtering is referred to~\cite{MSA:PFTtutorial2002}. Here, we aim to introduce notation and contextualise the discussion in subsequent sections.

Let $\{\mathbf{x}\}_{k=1,2,..}$ be the discrete-time Markov process
representing the collection of states and $\{\mathbf{z}\}_{k=1,2,..}$
be the sequence of measurements. $p(\mathbf{x}_k|\mathbf{x}_{k-1})$ is
the state transition probability and $p(\mathbf{z}_k|\mathbf{x}_k)$ is
the likelihood. Recursive Bayesian filtering is the solution to the
problem of using these models to process incoming data to obtain the
posterior probability density function,
$p(\mathbf{x}_k|\mathbf{z}_{1:k})$, where $\mathbf{z}_{1:k} =
\{\mathbf{z}_i, i=1, \ldots, k\}$ is the sequence of measurements up
to and including time $k$. $p(\mathbf{x}_k|\mathbf{z}_{1:k})$ is the
sufficient statistic used to calculate, for example, estimates of the
current state vector.

In a particle filter, the posterior is approximated using a set of $N$ random samples, where the $i$th sample is $\mathbf{x}^{i}_{k}$ and has a weight of $w_k^{i}$.



The weights are normalised such that
$\sum_{i=1}^{N}    w_k^i=1$. Estimates associated with the posterior at time $k$ can then be
approximated as:
\begin{equation}
\int f(\mathbf{x}_k)p(\mathbf{x}_{k}|\mathbf{z}_{1:k}) \approx \sum_{i=1}^{N}w_k^if(\mathbf{x}_{k}^i)\label{eq:weightedexp}
\end{equation}
where $f(.)$ is a function (e.g., $f(\mathbf{x}_k)=\mathbf{x}_k$ when calculating the mean). As the number of samples increases, the approximation becomes increasingly accurate. In fact, the variance of the estimate in~(\ref{eq:weightedexp}) can be shown to be upper-bounded by a quantity that is proportional to $\frac{1}{N}$.


\subsection{Sequential Importance Sampling}

Importance sampling~\cite{IS:1964} is a technique for approximating one pdf using weighted samples from another pdf. A Sequential Importance Sampler (SIS) involves applying importance sampling to the path\footnote{While the derivation involves consideration of a path, the resulting algorithm only needs to store the most recent state.} through the state-space, $x_{1:k}$. The samples up to time $k$ are also assumed to be generated by extending the samples of the path up to time $k-1$. This enables the weights in SIS to be derived as\cite{Doucet:2000:SMC}:
\begin{equation}
w_k^i \propto
\frac{
p(\mathbf{z}_k|\mathbf{x}_k^i)p(\mathbf{x}_k^i|\mathbf{x}_{k-1}^i)
}
{
q(\mathbf{x}_k^i|\mathbf{x}_{k-1}^i, \mathbf{z}_{k})
}
w_{k-1}^i\label{eq:sisweightupdate}
\end{equation}
where $q(\mathbf{x}_k|\mathbf{x}_{k-1}, \mathbf{z}_{k})$ is the {\em proposal distribution} used to generate $\mathbf{x}_k$ and where
\begin{equation}
w_1^i \propto
\frac{
p(\mathbf{z}_1|\mathbf{x}_1^i)p(\mathbf{x}_1^i)
}
{
q(\mathbf{x}_1^i)
}
\end{equation}
where $p(\mathbf{x}_1^i)$ and $q(\mathbf{x}_1^i)$ are distributions associated with the initial state and the initial distribution of samples (both at $k=1$).

Note that, when each measurement is received, SIS involves sampling particles from $q(\mathbf{x}_k|\mathbf{x}_{k-1}, \mathbf{z}_{k})$ and then updating their weights using~(\ref{eq:sisweightupdate}).

\subsection{Degeneracy Problem}

With the SIS algorithm, the variance of the importance weights can be proved to increase over time~\cite{Doucet:2000:SMC}\footnote{A good choice of proposal density can delay but not stop the effect\cite{Doucet:2000:SMC}.}. Empirically, this
results in {\em degeneracy}: all but one particle ends up having negligible normalised weights such that a single particle dominates the weighted average in~(\ref{eq:weightedexp}). A way to quantify this effect is to calculate the
effective sample size, $N_{eff}$, introduced in~\cite{Kong:1994} and
estimated as follows:
\begin{equation}
N_{eff} = \frac{1}{\sum_{i}^{N}(w_k^i)^2}
\end{equation}
where, since $0\leq w_k^i\leq 1$ and $\sum_{k=1}^{N} w_k^i=1$,
$1\leq N_{eff}\leq N$.

\subsection{Sequential Importance Resampling}\label{sec:sir}

$N_{eff}$ dropping below a threshold, $N_T$, indicates that estimates are likely to be inaccurate. The key to addressing this is to introduce {\em resampling}. The basic idea of resampling is to eliminate samples with low importance weights and
replicate samples with larger weights\footnote{This, of course, leads to a loss of diversity among the particles.}. While there are a number of variants of the resampling algorithm, they all consist of two core stages: calculating how many copies of each sample to generate; generating that number of copies of each sample. The different resampling variants differ in terms of how their calculate the number of copies to generate. We focus here on {\em minimum variance resampling} (also known as {\em systematic} resampling) which minimises the errors inevitably introduced by the resampling process (and is discussed in more detail in section~\ref{subsec:mvrs}). The use of resampling with SIS is often known as the Sampling Importance Resampling (SIR) filter and has been at the heart of particle filters since their invention~\cite{Gordon:1993,Kitagawa:1996,Isard:1998}.

\begin{small}
\begin{algorithm}[ht]
\begin{algorithmic}[1]
\STATE {\bf Function} sirFilter$(\text{ } p(x_0), \text{ } p(\mathbf{z}_\kappa|\mathbf{x}_\kappa), \text{ }p(\mathbf{x}_\kappa|\mathbf{x}_{\kappa-1}), \text{ }q(\mathbf{x}_\kappa|\mathbf{x}_{\kappa-1},\mathbf{z}_\kappa)\text{ })$\\
\STATE $\vartriangleright p(x_0) \text{: the initial prior}$
\STATE $\vartriangleright p(z_{\kappa}|x_\kappa)\text{: measurement model}$
\STATE $\vartriangleright p(x_{\kappa}|x_{\kappa-1})\text{: dynamic model}$
\STATE $\vartriangleright q(x_\kappa|x_{\kappa-1},z_{\kappa})  \text{: proposal}$

\STATE $\vartriangleright \text{Initialize the Particles}$
\STATE $\mathbf{x_0} \leftarrow \mathsf{drawSample}(p(x_0))$
\STATE $\mathbf{w_0} \leftarrow \frac{1}{N}$

\STATE $\vartriangleright \text{The time step loop}$

\FOR {$k=1$ \TO $T$}

  \STATE $\vartriangleright\!\text{Importance Sampling}$
  \STATE $\mathbf{x_k} \leftarrow \mathsf{drawSample}(q(\mathbf{x}_k|\mathbf{x}_{k-1},\mathbf{z}_{k}))$


  \STATE $ \vartriangleright\!\text{Calculate New Weights}$
  \STATE $\mathbf{w}^*_k\leftarrow \mathbf{w}_{k-1}\frac{p(\mathbf{z}_k|\mathbf{x}_k)p(\mathbf{x}_k|\mathbf{x}_{k-1})}{q(\mathbf{x}_k|\mathbf{x}_{k-1},\mathbf{z}_{k})}$


  \STATE $\vartriangleright \text{Normalise the Weights}$
  \STATE $\mathbf{w}_k\leftarrow\frac{\mathbf{w}^*_k}{\sum{\mathbf{w}^*_k}}$

  \STATE $\vartriangleright \text{Calculate Effective Sample Size}$
  \STATE $N_{eff} \leftarrow\frac{1}{ \sum{\mathbf{w_k^2}} }$


  \STATE $\vartriangleright \text{Perform Conditional Resampling}$
  \IF {$N_{eff} \leq N_t$}
      \STATE $\vartriangleright \text{The $\mathbf{m}$ represents the number of copies}$
      \STATE $\mathbf{m}\leftarrow\  \mathsf{minimumVarianceResampling}(\mathbf{w}_k)$
      \STATE $(\mathbf{m}, \mathbf{x}_k)\leftarrow\  \mathsf{quickSort}(\mathbf{m}, \mathbf{x}_k)$
      \STATE $\mathbf{x}_k \leftarrow \mathsf{redistribute}(\mathbf{m}, \mathbf{x}_k)$
  \ENDIF
  \STATE $\vartriangleright \text{Estimate the Mean (or any other quantities of interest)}$
  \STATE $\mathbf{\mu}_k\leftarrow\frac{\sum{\mathbf{x}_k}}{N}$
\ENDFOR
\STATE {\bf EndFunction}
\end{algorithmic}
\caption{SIR Filter --- Sequential (Vectorized) Version}
\label{alg:vectorized-sir-filter}
\end{algorithm}
\end{small}

Algorithm~\ref{alg:vectorized-sir-filter} shows pseudocode for the SIR filter. Note that the algorithm is expressed in vector notation, such that each vector operation implicitly comprises at least one {\em for} loop, and in terms of building blocks that operate on such vectors. The algorithm
relies on a number of functions, which are covered in detail later in this paper. Briefly these functions include:
\begin{itemize}
\item {\tt ($a$)$\leftarrow$ drawSample($q(.)$)} draws samples from the supplied distribution, $q(.)$;
\item {\tt ($m$) $\leftarrow$ minimumVarianceResampling($w$)} determines the number of times
each particle needs to be replicated. The function takes the particles' weights, $w$, as
input.
\item {\tt ($m$,$x$) $\leftarrow$  quickSort($m$,$x$)} calculates the permutation that would sort vector $m$, and applies this permutation to both inputs. While this sort is not necessary with a single processor implementation, we will exploit the fact that the output has been sorted in section~\ref{sec:improvement}.
\item{\tt $x' \leftarrow$ redistribute($m$,$x$)} returns the new population of particles, $x'$, where $m$, as mentioned previously, defines the number of replications of each of the old population of particles, $x$.
\end{itemize}

\subsection{Parallel Particle Filtering}

The bulk of the operations comprising the particle filter (as described in Algorithm~\ref{alg:vectorized-sir-filter}) are readily parallelised. However, it is resampling (the redistribution process in particular) that complicates parallel implementation of particle filters.

The complications primarily arise because, if each of multiple processors are considering subsets of the particles, the data transfers that the redistribution process demands are data-dependent. It is therefore non-trivial to implement a particle filter in a way that the run-time is not data-dependent. A similar problem has been encountered with sorting algorithms\footnote{For instance, although Quicksort~\cite{Hoare:1961} can be parallelised, the load
distributions across the processors is dependent on the pivots used and the run-time will therefore be data-dependent.}. In the subsequent sections of this paper, we describe how to implement the components of the particle filter in a way that run-time in not data-dependent, but deterministic.
\section{Parallel Instantiations of the Algorithmic Components of Particle Filtering}
\label{sec:components}

Prior to mapping the particle filter algorithm on to a MapReduce form,
it is essential to understand how the operations used by a particle filter can be implemented in a fully distributed form. While a more detailed discussion of these operations (and others) can be found in~\cite{Blelloch:1990}, we now discuss each of the
operations that constitute the algorithm described in algorithm~\ref{alg:vectorized-sir-filter}. We summarise these operations and the
associated complexities in table~\ref{tab:pf-operation-complexities}, both for the fundamental building blocks and some of the algorithmic components that can be built from those components. Our focus is on implementations with a time-complexity that is as fast as possible in terms of its dependence on $N$, the number of data.

\begin {table*}[!ht]
\centering
    \caption{Complexities of Various Algorithmic Components of the Particle Filter.}
\label{tab:pf-operation-complexities}
\begin{tabular}{llrr}
\toprule
Section & Algorithmic Component  & Time & Space \\
\midrule
\ref{sec:elemental} & Element-wise operations & 
$\bigo{1}$ & $\bigo{N}$\\
\ref{sec:rotation}& Rotation   &
$\bigo{1}$ 			& $\bigo{N}$ \\
\ref{sec:sum}& Sum/Max/Min & 
$\bigo{\log N}$ 		& $\bigo{N}$ \\
\ref{subsec:scan}&Cumulative Sum    		& 
$\bigo{\log N}$ 		& $\bigo{N}$ \\
\midrule
\ref{sec:norm} &Normalising the Weights 			& 
$\bigo{\log N}$ 		& $\bigo{N}$ \\
\ref{subsec:mvrs} &Minimum Variance Resampling 		& 
$\bigo{\log N}$ 		& $\bigo{N}$ \\
\ref{subsec:sorting} & (Bitonic) Sort & 
$\bigo{(\log N)^2}$  	        & $\bigo{N}$ \\
\ref{subsec:redistribution}& Redistribution from~\cite{Maskell:SIMDPF2006}				&
$\bigo{(\log N)^3}$ 	        &
$\bigo{N}$ \\
\ref{sec:improvement}& Improved Redistribution				&
$\bigo{(\log N)^2}$ 	        &
$\bigo{N}$ \\
\bottomrule
\end{tabular}
\end{table*}


\subsection{Element-wise Operations}\label{sec:elemental}

Perhaps the simplest type of operation to implement in parallel involves applying an
element-wise operation\footnote{Such operations are an example of `embarrassingly parallel' operations that are arguably trivial to parallelise.}. Given a function $f$ and a vector
$\mathbf{v}$, the element-wise operation $f\mapsto\mathbf{v}$ applies
the function $f$ on every element of the vector such that:
$$
f\mapsto\mathbf{v} = [f(v_1), f(v_2), \ldots, f(v_{N})]
$$

In our case, normalizing the weights is an example of an element-wise
operation. Another example is a vector of \emph{if} operations, $\mathsf{Vif}(\mathbf{a},\mathbf{b},\mathbf{c})$ where the $i$th element in the output is $b_i$ if $a_i$ is true and $c_i$ otherwise.

It should be evident that operations that involve two inputs and a single output (e.g., element-wise sum or difference) are similarly easy to implement in parallel.

\subsection{Rotation}\label{sec:rotation}

Another operation that we will use involves rotating (with wrapping (i.e., cyclic shift) or without) the elements of a vector by a given distance, $\delta$, such that if the input is $\mathbf{a}$ and the output is $\mathbf{b}$, after the rotation, we have $b(\mod(i+\delta,N))=a(i)$ where $\mod(x,y)$ is $x$ modulus $y$. Once again, this algorithmic component is readily parallelised.

We will also use partial rotations such that we have a vector of distances, $\Delta$, and not a single `global' distance, $\delta$. This vector, $\Delta$, has $N'<N$ elements where $N'$ is a power of two. The rotations are then implemented locally to each set of $M=\frac{N}{N'}$ elements. For example if the $j$th element of $\Delta$ is $\delta_j$ then $b\left(\left(j-1\right)\times M+\mod\left(i+\delta_j,M\right)\right)=a\left(\left(j-1\right)\times M+i\right)$ for $1\leq i\leq M$.

\subsection{Sum, Max and other Commutative Operations}\label{sec:sum}

To calculate a sum of a vector of numbers, we can use an `adder-tree'. The numbers are associated with the leaves of the tree. By recursing up the tree, the sum of pairs of numbers can be calculated (in parallel across all pairs). The sum of all pairs of pairs of numbers can then be calculated (in parallel across all pairs of pairs). This is exemplified in figure~\ref{fig:cumsum_example}(a-c). This process can repeat until we reach the root node of the tree and calculate the sum of all the numbers by summing the sum of the two halves of the data. See figure~\ref{fig:cumsum_example}(d).

In fact, as has been known since the development of the infamous Array Programming Language (APL)~\cite{Iverson:1962},
this same approach can be used for any binary operation, $\oplus$, that is commutative such that:
\begin{equation}
((a\oplus b)\oplus c)\oplus d=(a\oplus b)\oplus(c\oplus d)
\end{equation}

Relevant examples of operations which can be calculated in this way include the sum but also the maximum (and minimum) and first non-zero element of a set of numbers (which we will denote $\mathsf{First}(.)$ in, for example, algorithms~\ref{alg:redistribute-old} and~\ref{alg:shift}). For such operations, with $N$ processors processing $N$ data and a binary tree, the time-complexity is the depth of the tree, i.e., $\log_2 N$. 

As should be evident, an upside-down version of the same tree can be used to implement an $\mathsf{Expand}(a)$ operation, which involves making all elements of a vector equal to the single value of $a$.

\subsection{Cumulative Sum}
\label{subsec:scan}

While the ability to use a tree to calculate a sum efficiently is well known, the ability to use a closely related approach to calculate a cumulative sum\footnote{Note that the cumulative sum is sometimes referred to as a prefix sum: there is no difference between a prefix sum and a cumulative sum.} efficiently appears to be less well known by researchers working on particle filters. Of course, a na\"{i}ve implementation involves computing the cumulative sum by simply adding each element of the input to the previous element of the output. Such an approach would have a run-time of $N$.
However, a more-efficient approach has existed since the development of APL if not for longer\footnote{APL describes an approach to calculating a sum, maximum or minimum as \emph{reduction} operations. The approach to calculating a cumulative sum is described as a \emph{scan} operation and can be used to calculate, for example, cumulative maximums and minimums. Scan operations take a binary operator $\oplus$ and an $N$-element
vector $\mathbf{a}=[a_1, a_2, \ldots, a_{N}]$, and return an $N$-element vector $\mathbf{a}_{\oplus}=[a_1, (a_1\oplus a_2), \ldots,
  (a_1\oplus a_2 \oplus a_3 \oplus \ldots \oplus a_{N})]$. However, here we are only concerned with cumulative sums.}.

\begin{figure}
\begin{center}
\includegraphics[width=1\textwidth]{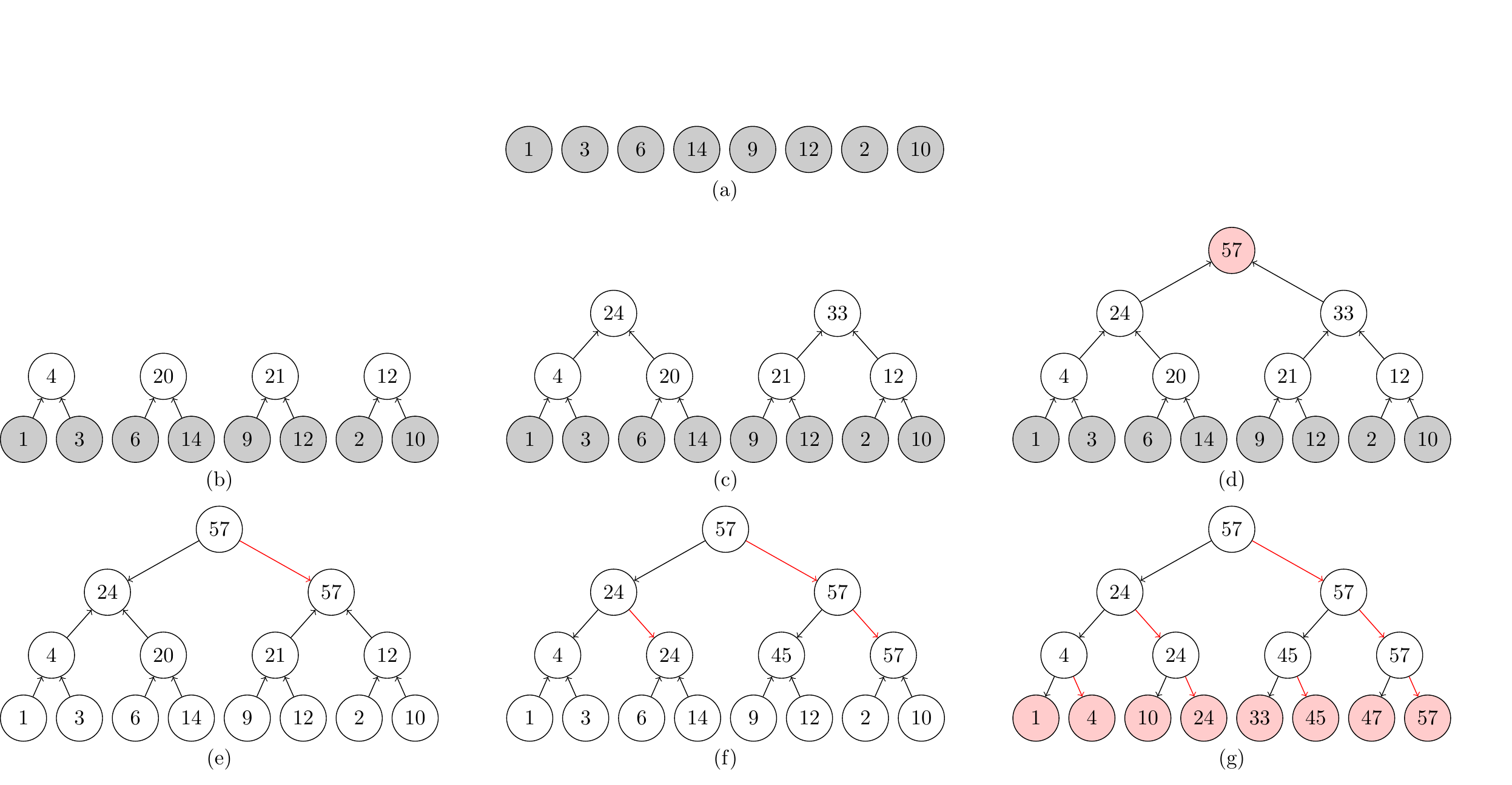}
\caption{Example of cumulative sum for N=8 numbers. Subfigures (a)-(d) describe the sum computation, while the remaining balanced binary trees shown in subfigures (e)-(g) describe how the backward pass culminates in calculation of the cumulative sum of the given sequence.}
\label{fig:cumsum_example}
\end{center}
\end{figure}

To ensure the reader has some intuition as to how this could be possible, the key idea is to exploit the partial sums that are calculated in an adder-tree and to express each element of the cumulative sum as a sum of these (efficiently calculated) partial sums. The process that exploits this insight then involves a second tree in which the values at every level are propagated to the
level below, replacing the values that were calculated in the adder-tree. More specifically, in the downward propagation, the value at each parent node is propagated to its right child and to its left child. The new value for the left child is the difference between the value at the parent node and the value at the right child node (as calculated in the adder-tree). The new value for the right child is just the same value as the parent node. See figures~\ref{fig:cumsum_example}(e)-(g) for an example.

With this forward and backward pass of the tree, we can obtain the
cumulative sum in $2\log_2 N$ steps.

\subsection{Normalising the Weights}\label{sec:norm}
Normalising the weights is an example of an operation that can be implemented using the building blocks described to this point. The sum is calculated using an adder-tree (as described in section~\ref{sec:sum}), distributed to all the data (as also described in section~\ref{sec:sum}) and an element-wise divide (see section~\ref{sec:elemental}) used to calculate the normalised weights.

\subsection{Minimum Variance Resampling}
\label{subsec:mvrs}

As explained in section~\ref{sec:sir}, resampling involves determining the number of copies of each particle that are needed. We specifically describe minimum
variance resampling, for which the number of copies of the $i$th particle is:
\begin{equation}
m_i = \lfloor C_{i}\times N\rfloor-\lceil C_{i-1}\times N\rceil+1\label{eq:mvrs}
\end{equation}
where $\lceil x\rceil$ and $\lfloor x \rfloor$ are respectively the ceiling\footnote{The ceiling of $x$ is the smallest integer larger than or equal to $x$.} and the floor\footnote{The floor of $x$ is the largest integer smaller than or equal to $x$.} of $x$, where
\begin{equation}
C_i = \sum_{j=1}^i w_i+\epsilon\label{eq:cumulativeweights}
\end{equation}
is the cumulative sum and where $\epsilon\sim\left[0,\frac{1}{N}\right]$ and $C_0=0$.

(\ref{eq:mvrs}) uses only element-wise operations (as described in section~\ref{sec:elemental}) and a rotation (by a single element and as described in section~\ref{sec:rotation}). (\ref{eq:cumulativeweights}) involves a cumulative sum (as described in section~\ref{subsec:scan}) and an addition (as described in section~\ref{sec:elemental}). Thus, the building blocks described to this point can be used to implement~(\ref{eq:mvrs}) and~(\ref{eq:cumulativeweights}).








\subsection{Sorting}
\label{subsec:sorting}

Quicksort~\cite{Hoare:1961} is well known and has an
average time complexity of $\bigo{N\log_2 N}$. However, we focus on the bitonic sort algorithm~\cite{Batcher:1968}, which has a
time complexity of $\bigo{\left(\log_2 N\right)^2}$ and a spatial complexity of $\bigo{N}$. The main reason for this choice is that we want to guarantee the time taken to perform sorting. While it is possible to parallelise quicksort, the ability to do so is data dependent. In contrast, bitonic sort has deterministic time complexity (with a balanced load across (up to) $N$ processors).

At the fundamental level, a {\em bitonic sequence} forms the basis for the bitonic sort. A sequence $\mathbf{a}=[a_1, a_2, \ldots, a_N]$ is a
bitonic sequence if $a_1\leq a_2\leq \ldots \leq a_k \geq \ldots \geq a_N $ for some $k$, $1 \leq k \leq N$ or if this condition holds for any rotation of $\mathbf{a}$.

To try to provide some intuition as to how the algorithm works, note that at a certain point it the algorithm, we have $N$ data in a bitonic sequence. The first `half' of the data are sorted in ascending order and the second half are sorted in descending order\footnote{A similar argument works if the first half are sorted in descending order and the second half are sorted in ascending order.}. Consider the $i$th element in the first half and the $i$th element in the second half. There are $\frac{N}{2}-1$ data between these two elements. They must all be larger than the smallest of the two elements which the data are between. There must therefore be at least $\frac{N}{2}$ data that are larger than the smallest of the two elements. This smallest element must therefore be one of the lowest $\frac{N}{2}$ data (it cannot be one of the largest $\frac{N}{2}$ data if there are at least $\frac{N}{2}$ data larger than it). An upside-down version of the same argument makes clear that the largest of these two elements must be one of the largest $\frac{N}{2}$ data. Finally, it also transpires that after this operation, the first $\frac{N}{2}$ data are a bitonic sequence and the second $\frac{N}{2}$ data are a bitonic sequence. Thus, given a bitonic sequence, by comparing all pairs of data that are a distance of $\frac{N}{2}$ apart and swapping the points if needed, we can ensure all the larger elements are in the first $\frac{N}{2}$ data, which forms a bitonic sequence, and all the smaller elements are in the second $\frac{N}{2}$ data, which also forms a bitonic sequence. We can then apply the same comparison structure on each of the two bitonic (smaller) sequences. This process can be applied recursively until pairs of points are compared and the data are sorted.

This process is known as the `bitonic merge' and requires $\bigo{\log_2 N}$ steps (with $\bigo{N}$ spatial complexity) to convert a bitonic sequence into a sorted sequence. To generate the bitonic sequence needed from arbitrary input data\footnote{This process is sometimes known as `bitonic build'.}, we apply bitonic sort to put the first $\frac{N}{2}$ input data into ascending order and apply bitonic sort again to put the second $\frac{N}{2}$ input data into descending order. Analysis of this recursive use of bitonic sort gives rise to bitonic sort requiring $\frac{n^2 - n}{2}$ iterations where $n=\log_2 N$ and, at every step, the algorithm performs $\frac{N}{2}$ comparisons. Each comparison involves comparing two data and swapping them according to a criterion that is defined by the position of the comparison in the network (and can be implemented using the building blocks described in sections~\ref{sec:elemental} and~\ref{sec:rotation}).

\begin{figure}[!ht]
    \centering
    \includegraphics[width=0.6\textwidth]{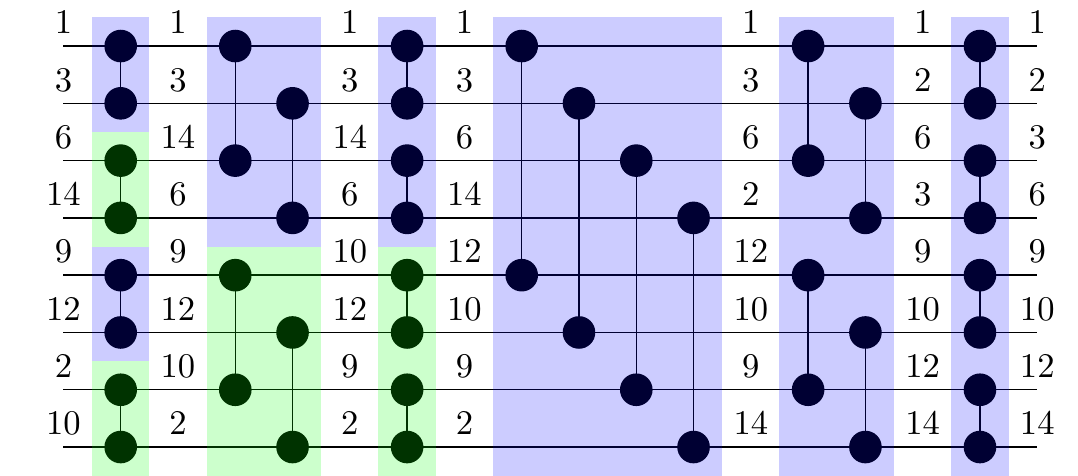}
\caption{Example of bitonic sort using eight numbers. Each horizontal wire corresponds to a core. The blue color denotes that the larger value will be stored at the lower wire after the comparison, while the green color the opposite.}
\label{fig:bsort}
\end{figure}

An example of bitonic sort with eight numbers is provided in figure~\ref{fig:bsort}.

\subsection{Redistribution}

\subsubsection{Original Version from~\cite{Maskell:SIMDPF2006}}
\label{subsec:redistribution}

The redistribution algorithm takes two inputs,
the old population of particles $\mathbf{x}$, and the number of copies
$\mathbf{m}$, and produces the  new
population of particles, $\mathbf{x}^*$, as the output.

\begin{figure}
\begin{center}
\includegraphics[width=0.75\textwidth]{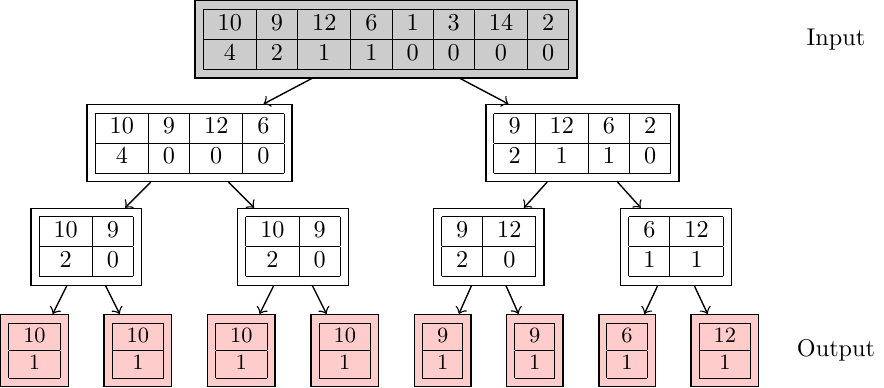}
\caption{An example of the redistribution for $\mathbf{x}=[10, 9, 12,
    6, 1, 3, 14, 2]$ and $\mathbf{m}=[4, 2, 1, 1, 0, 0, 0, 0]$. The
  first and the second rows represent the old population of particles
  and the number of copies needed to generate the new population,
  $\mathbf{x}^* = [10, 10, 10, 10, 9, 9, 6, 12] $.}
\label{fig:redistribute-example}
\end{center}
\end{figure}

In~\cite{Maskell:SIMDPF2006}, a divide-and-conquer algorithm was described for implementing the redistribute. The procedure involves sorting the particles in decreasing order of the number of copies. With $N$ data, the sum of the elements of $\mathbf{m}$ must be $N$. The approach is then to divide the data into two smaller datasets, each of which has $\frac{N}{2}$ elements and is such that the corresponding elements of $\mathbf{m}$ are sorted and sum to $\frac{N}{2}$. This can be achieved by finding the \emph{pivot}, which we define as leftmost element in $\mathbf{m}$ for which the associated value of the cumulative sum is $\frac{N}{2}$ or greater. In general, the pivot needs to be split into two constituent parts such that the two smaller datasets can both sum to $\frac{N}{2}$. We refer to these two parts as the left-pivot and right-pivot. The data to the left of the pivot and including the left-pivot can be used to produce one of the two smaller datasets. The right-pivot and the data to the right of the pivot can be used to produce the other of the two smaller datasets. Both smaller datasets are then sorted\footnote{The first dataset is actually already sorted, but the second dataset is, in general, not sorted.} such that they are in decreasing order of $\mathbf{m}$. Note that there is a special case that occurs when the value of the right-pivot is zero: the rotation needed is one less than otherwise in this case. It can be intuitive to think of this procedure as operating on a tree. Applying the procedure recursively down the tree, until the leaf nodes are encountered, results in the redistribute completing. See figure~\ref{fig:redistribute-example} for an illustrative example of this procedure.

The procedure can be described using element-wise operations (see section~\ref{sec:elemental}), sum (see section~\ref{sec:sum}), cumulative sum (see section~\ref{subsec:scan}), rotations (see section~\ref{sec:rotation}) and sort (see section~\ref{subsec:sorting}). Algorithm~\ref{alg:redistribute-old} provides a description of this algorithm. Note that the description makes use of three functions ($\mathsf{LeftHalf}(.)$, $\mathsf{RightHalf}(.)$ and $\mathsf{Combine}(.)$) which are included to aid exposition (and actually have zero computational cost). Also note that the implementation is described in a way that involves recursion. It is possible to `unwrap' the recursive implementation such that all operations (at all stages in the tree) are implemented on datasets of the same size (a size of $N$). Doing so is conceptually straightforward though the bookkeeping required is non-trivial.

\begin{small}
\begin{algorithm}[ht]
\begin{algorithmic}[1]
\STATE {\bf Function} $\mathbf{x} = \mathsf{Redistribute}(\mathbf{m}, \mathbf{x})$\\
\STATE $\vartriangleright \mathbf{m}$: Number of copies (sorted in descending order)\\
\STATE $\vartriangleright \mathbf{x}$: Particles \\

\IF {$\mathsf{Length}(\mathbf{m})>1$}
\STATE $\vartriangleright$ Calculate Cumulative Sum
\STATE $\mathbf{c} \leftarrow \mathsf{CumSum}(\mathbf{m})$

\STATE $\vartriangleright$ Identify Pivot
\STATE $i_p \leftarrow \mathsf{First}(\mathbf{c}\ge\frac{N}{2})$
\STATE $\mathbf{p} \leftarrow \mathsf{Expand}(i_p)$

\STATE $\vartriangleright$ Calculate Left-Pivot and Right-Pivot
\STATE $\vartriangleright$ $\mathbf{i}$ simply indexes the elements of $\mathbf{m}$ and $\mathbf{0}$ is a vector of zeros
\STATE $\mathbf{lp} \leftarrow \mathsf{Vif}(\mathbf{i}=\mathbf{p},\mathbf{c}-\frac{N}{2},\mathbf{0})$
\STATE $\mathbf{rp} \leftarrow \mathsf{Vif}(\mathbf{i}=\mathbf{p},\frac{N}{2}-\mathsf{Rotate}(\mathbf{c},1),\mathbf{0})$

\STATE $\vartriangleright$ Generate Smaller Datasets
\STATE $\mathbf{l} \leftarrow \mathsf{LeftHalf}(\mathsf{Vif}(\mathbf{i}<\mathbf{p},\mathbf{m},\mathbf{lp}))$
\STATE $\mathbf{lx} \leftarrow \mathsf{LeftHalf}(\mathbf{x})$
\STATE $\mathbf{r} \leftarrow \mathsf{Vif}(\mathbf{i}>\mathbf{p},\mathbf{m},\mathbf{rp})$

\STATE $\vartriangleright$ Calculate Rotation of $\mathbf{r}$
\STATE $inc \leftarrow \mathsf{Sum}(\mathsf{Vif}(\mathbf{c}=\frac{N}{2},\mathbf{1},\mathbf{0}))$
\STATE $\mathbf{r} \leftarrow \mathsf{RightHalf}(\mathsf{Rotate}(\mathbf{r},i_p+inc))$
\STATE $\mathbf{rx} \leftarrow \mathsf{RightHalf}(\mathsf{Rotate}(\mathbf{x},i_p+inc))$

\STATE $\vartriangleright$ Sort Right Half
\STATE $\mathbf{r} \leftarrow \mathsf{Sort}(\mathbf{r})$

\STATE $\vartriangleright$ Divide-and-conquer
\STATE $\mathbf{lx} \leftarrow  \mathsf{Redistribute}(\mathbf{l},\mathbf{lx})$
\STATE $\mathbf{rx} \leftarrow  \mathsf{Redistribute}(\mathbf{r},\mathbf{rx})$

\STATE $\vartriangleright$ Combine Outputs
\STATE $\mathbf{x} \leftarrow \mathsf{Combine}(\mathbf{lx},\mathbf{rx})$
\ENDIF

\STATE {\bf EndFunction}
\end{algorithmic}
\caption{Redistribute: $\bigo{(\log_2 N)^3}$ implementation.}
\label{alg:redistribute-old}
\end{algorithm}
\end{small}

The time complexity of this redistribution algorithm $\bigo{(\log_2 N)^3}$ in parallel with $N$ processors since a (bitonic) sort (with complexity of $\bigo{(\log_2 N)^2}$) is used at each stage in the divide-and-conquer. Note that this contradicts the (erroneous) claim in~\cite{Maskell:SIMDPF2006} that the time complexity of this algorithm is $\bigo{(\log_2 N)^2}$.

\subsubsection{Improved Redistribution}
\label{sec:improvement}

The redistribution algorithm described in section~\ref{subsec:redistribution} is a
divide-and-conquer algorithm that ensures that, at each node in the tree, $\mathbf{m}$ sums to its length, $N$, and is sorted. The sorting is sufficient to ensure that rotation can be used to replace some of the (rightmost) zeros with the (rightmost) non-zero elements of $\mathbf{m}$ that sum to $\frac{N}{2}$.

Here we exploit the observation that it is possible to define an alternative divide-and-conquer strategy. More specifically, we ensure that, at each node in the tree, $\mathbf{m}$ sums to its length, $N$, and has all its non-zero values to the left of all values that are zero. Since such a sequence only has trailing zeros, we call such a sequence an All-Trailing-Zeros (ATZ) sequence\footnote{We suspect such a sequence may have a name in a literature we do not currently have sight of. However, here we simply adopt an intuitive name for ease of exposition.}. While a sort is sufficient to generate an ATZ sequence, it is easier, as we will demonstrate shortly, to generate an ATZ sequence than it is to generate a sorted sequence.

The new algorithm, at each node in the tree, starts with $\mathbf{m}$, which sums to its length, $N$, and is an ATZ sequence. To proceed, as previously, we find the pivot (as defined in section~\ref{subsec:redistribution}). As previously, the data to the left of the pivot and the left-pivot can be used to produce one of the two smaller datasets. However, in contrast to the approach described in section~\ref{subsec:redistribution}, we can simply use the right-pivot and the data to the right of the pivot to generate the second smaller dataset (without any need for sort). Both these smaller datasets then sum to $\frac{N}{2}$ and are ATZ sequences. Note that, as with the approach described in section~\ref{subsec:redistribution}, there is a special case that occurs when the value of the right-pivot is zero.

To initiate the algorithm, we need to generate an ATZ sequence. To achieve this, we propose to use (bitonic) sort (once). After this initial sort, the procedure can be described using element-wise operations (see section~\ref{sec:elemental}), sum (see section~\ref{sec:sum}), cumulative sum (see section~\ref{subsec:scan}) and rotations (see section~\ref{sec:rotation}). We emphasise that there is no need for a sort after the initial generation of an ATZ sequence. As a result, while the algorithm described in section~\ref{subsec:redistribution} has time-complexity of $\bigo{(\log_2 N)^3}$, the algorithm described in this section has time-complexity of $\bigo{(\log_2 N)^2}$. To aid understanding algorithm~\ref{alg:shift} provides a description of this algorithm. Note the very strong similarity to algorithm~\ref{alg:redistribute-old} and that, once again, it is possible to `unwrap' the recursive implementation albeit with some non-trivial bookkeeping.

\begin{small}
\begin{algorithm}[ht]
\begin{algorithmic}[1]
\STATE {\bf Function} $\mathbf{x} = \mathsf{Redistribute}(\mathbf{m}, \mathbf{x})$\\
\STATE $\vartriangleright \mathbf{m}$: Number of copies (in an ATZ sequence)\\
\STATE $\vartriangleright \mathbf{x}$: Particles \\

\IF {$\mathsf{Length}(\mathbf{m})>1$}
\STATE $\vartriangleright$ Calculate Cumulative Sum
\STATE $\mathbf{c} \leftarrow \mathsf{CumSum}(\mathbf{m})$

\STATE $\vartriangleright$ Identify Pivot
\STATE $i_p \leftarrow \mathsf{First}(\mathbf{c}\ge\frac{N}{2}))$
\STATE $\mathbf{p} \leftarrow \mathsf{Expand}(i_p)$

\STATE $\vartriangleright$ Calculate Left-Pivot and Right-Pivot
\STATE $\vartriangleright$ $\mathbf{i}$ simply indexes the elements of $\mathbf{m}$ and $\mathbf{0}$ is a vector of zeros
\STATE $\mathbf{lp} \leftarrow \mathsf{Vif}(\mathbf{i}=\mathbf{p},\mathbf{c}-\frac{N}{2},\mathbf{0})$
\STATE $\mathbf{rp} \leftarrow \mathsf{Vif}(\mathbf{i}=\mathbf{p},\frac{N}{2}-\mathsf{Rotate}(\mathbf{c},1),\mathbf{0})$

\STATE $\vartriangleright$ Generate Smaller Datasets
\STATE $\mathbf{l} \leftarrow \mathsf{LeftHalf}(\mathsf{Vif}(\mathbf{i}<\mathbf{p},\mathbf{m},\mathbf{lp}))$
\STATE $\mathbf{lx} \leftarrow \mathsf{LeftHalf}(\mathbf{x})$
\STATE $\mathbf{r} \leftarrow \mathsf{Vif}(\mathbf{i}>\mathbf{p},\mathbf{m},\mathbf{rp})$

\STATE $\vartriangleright$ Calculate Rotation of $\mathbf{r}$
\STATE $inc \leftarrow \mathsf{Sum}(\mathsf{Vif}(\mathbf{c}=\frac{N}{2},\mathbf{1},\mathbf{0}))$
\STATE $\mathbf{r} \leftarrow \mathsf{RightHalf}(\mathsf{Rotate}(\mathbf{r},i_p+inc))$
\STATE $\mathbf{rx} \leftarrow \mathsf{RightHalf}(\mathsf{Rotate}(\mathbf{x},i_p+inc))$

\STATE $\vartriangleright$ Divide-and-conquer
\STATE $\mathbf{lx} \leftarrow  \mathsf{Redistribute}(\mathbf{l},\mathbf{lx})$
\STATE $\mathbf{rx} \leftarrow  \mathsf{Redistribute}(\mathbf{r},\mathbf{rx})$

\STATE $\vartriangleright$ Combine Outputs
\STATE $\mathbf{x} \leftarrow \mathsf{Combine}(\mathbf{lx},\mathbf{rx})$
\ENDIF

\STATE {\bf EndFunction}
\end{algorithmic}
\caption{Redistribute: $\bigo{(\log_2 N)^2}$ implementation.}
\label{alg:shift}
\end{algorithm}
\end{small}

\section{Mapping Particle Filtering into MapReduce}
\label{sec:mrpf}

The descriptions provided in the section~\ref{sec:components} describe distributed operations that can manipulate vectors (albeit after some unwrapping of the recursive descriptions).

As discussed in the section~\ref{sec:mapreduce}, the fundamental
notion of MapReduce is the processing of (key, value) pairs.  In the
context of particle filtering, none of the properties of the particles
(weight or state) qualifies to be a key. However, we can give each particle a unique index and use this index as the key, such that we think of the particles as being a set $\left\{i, x_i, w_i\right\}$ where  $i\in\{1, \ldots, N\}$ and, as previously, where $N$
is the number of particles, $x_i$ is the state and $w_i$ is the
corresponding weight of the $i$th particle.

\section{Evaluation}
\label{sec:evaluation}

\begin{table}
\begin{center}
    \begin{tabular}{|l|l|l|}
    \hline
    {\bf Details}          & {\bf Single Node System} & {\bf Multi-Node System} ~ \\
    \hline
    \hline
    Name				   & Platform~1 	& Platform~2 ~\\
    Number of Nodes        & 1    & 28                  ~ \\
    Hardware Cores         & 16   & 512                 ~ \\
    Operating System       & Linux & IBM Unix           ~ \\
    Primary Memory         & 16GB  & 384GB              ~ \\
    Spark Version          & 1.6.2  & 1.4.1\\
    Hadoop Version         & 2.7.2  & 2.7.1 \\
    \hline
    \end{tabular}
 \caption{Details of the Experimental Platform used for Evaluation.}
 \label{tab:sysdetails}
 \end{center}
\end{table}

We performed extensive evaluation of our algorithm on two
different systems. We provide the details of these systems in
table~\ref{tab:sysdetails}. The evaluation process included the
algorithms outlined in the section~\ref{sec:components}
on the two key frameworks that support MapReduce and which were mentioned in section~\ref{sec:bigdata}: Hadoop and Spark. We used the standard estimation
problem (involving a scalar state and a computationally inexpensive proposal, likelihood and dynamic model) that is widely used in the particle filtering
community~\cite{MSA:PFTtutorial2002}. We perceive this scenario emphasises the need for efficient resampling: were, as is often the case, the likelihood, dynamics and proposal were computationally demanding, the relative merits of different resampling schemes would be less apparent. Our evaluation focused on specific aspects of the implementation, which are described as follows:
\begin{enumerate}
\item We start, in section~\ref{sec:worst}, by providing evidence that, in contrast to a na\"{i}ve implementation, the particle filter we have developed can exploit multi-core architectures while having deterministic run-time.

\item In section~\ref{sec:profile}, as a precursor to a detailed evaluation and analysis, we analyse the overall profile of the particle filtering
  algorithm for implementations on a single core, using Hadoop and using Spark.

\item Then, in section~\ref{sec:hadoopspark}, for both the Spark and Hadoop implementations, we compare the performance of our new algorithms relative to a single mapper and a single reducer. In doing so, we not only compare the overall performance,
  but we also compare the fundamental building blocks of the particle filtering
  algorithm. This section provides a thorough understanding of these algorithms' performance on two key frameworks that support MapReduce.

\item Given that the Spark implementation (unsurprisingly) outperforms the Hadoop implementation, we then focus on the Spark implementation. In section~\ref{sec:evlaution:multicore} we then compare the two versions of the redistribution algorithm described in sections~\ref{subsec:redistribution} and~\ref{sec:improvement} as a function of the numbers of particles and cores. The intent is that this detailed comparison provides insight into the performance that is achievable using the original and proposed variants of the redistribution algorithm.

\item Finally, in section~\ref{sec:speedup}, we perform a detailed analysis on the speedup and scalability of the redistribution and the overall particle filter.
\end{enumerate}

In performing these evaluations, a basic parameter that we
found useful in assessing the algorithmic performance is the
capability to process large amount of data, which directly translates
to the number particles that can be processed per unit time, the number
of Particles Processed per Second (PPS).

\subsection{Worst Case Runtime Performance}\label{sec:worst}

\subsubsection{Baseline Redistribution Algorithm}

We will compare performance against a na\"{i}ve baseline implementation of the redistribution component. This implementation involves calculation (in parallel) of a cumulative sum of the number of copies. Once this cumulative sum is calculated (and each element of the sum communicated to be processed along with its neighbour), for each particle in the old population, we know the first and last indices of particles in the new population that will be copies of this particle in the old population. Then, by performing a loop across the particles in the old population, we can populate the new generation of particles.

Note that this algorithm, when running across multiple cores, can be expected to have a runtime complexity that is dependent on the data. To help make this clear, consider the worst case where the redistribution involves making $N$ copies of the $i$th particle (and zero copies of all other particles). In this case, only one core will actually be populating the new generation of particles.

\subsubsection{Runtime Performance}

\begin{figure}
\begin{center}
\begin{tabular}{|c|}
\hline
    \includegraphics[width=0.44\textwidth]{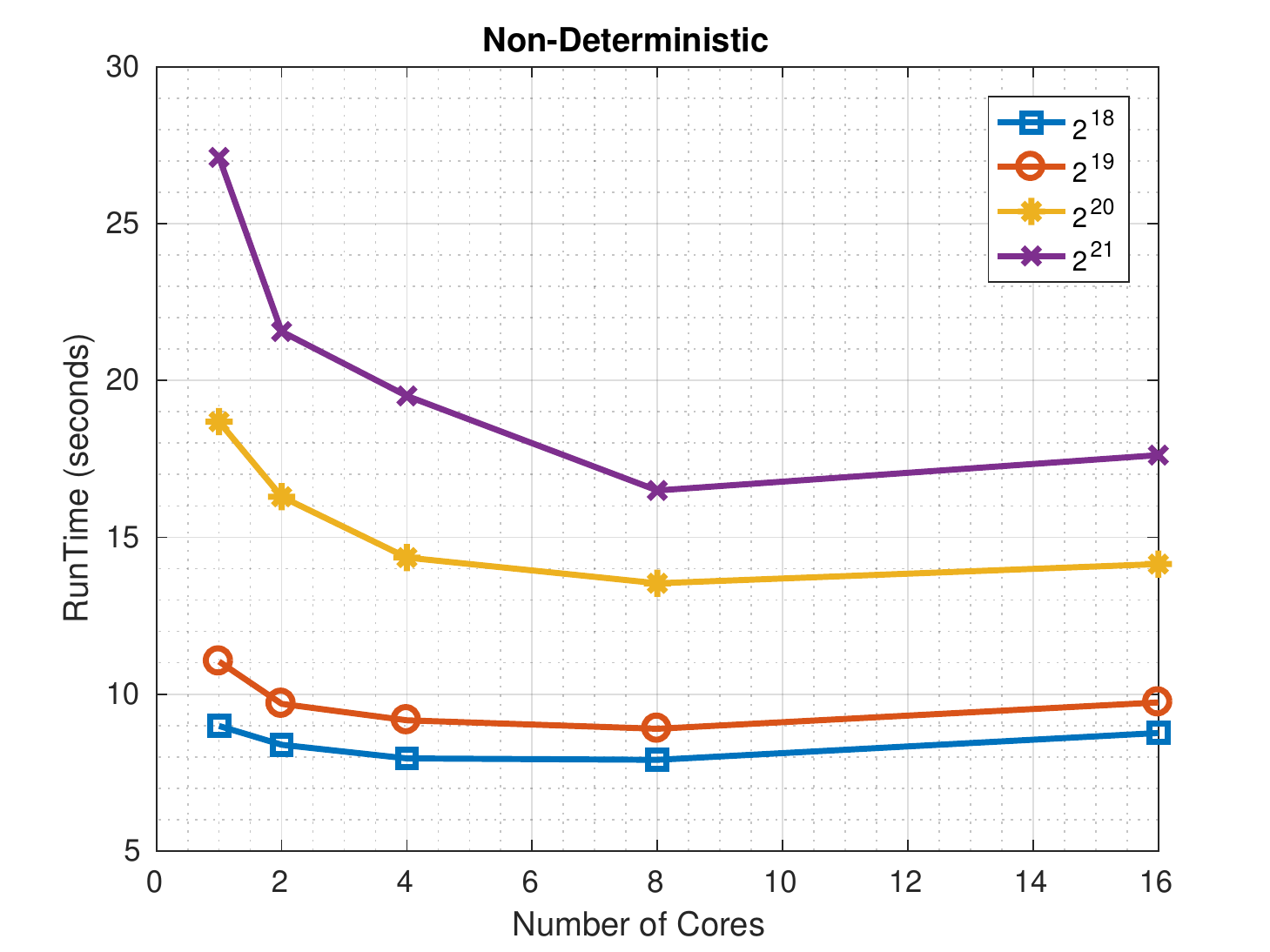} \\
    (a) Na\"{i}ve implementation \\
\hline
\end{tabular}
\quad
\begin{tabular}{|c|}
\hline
    \includegraphics[width=0.44\textwidth]{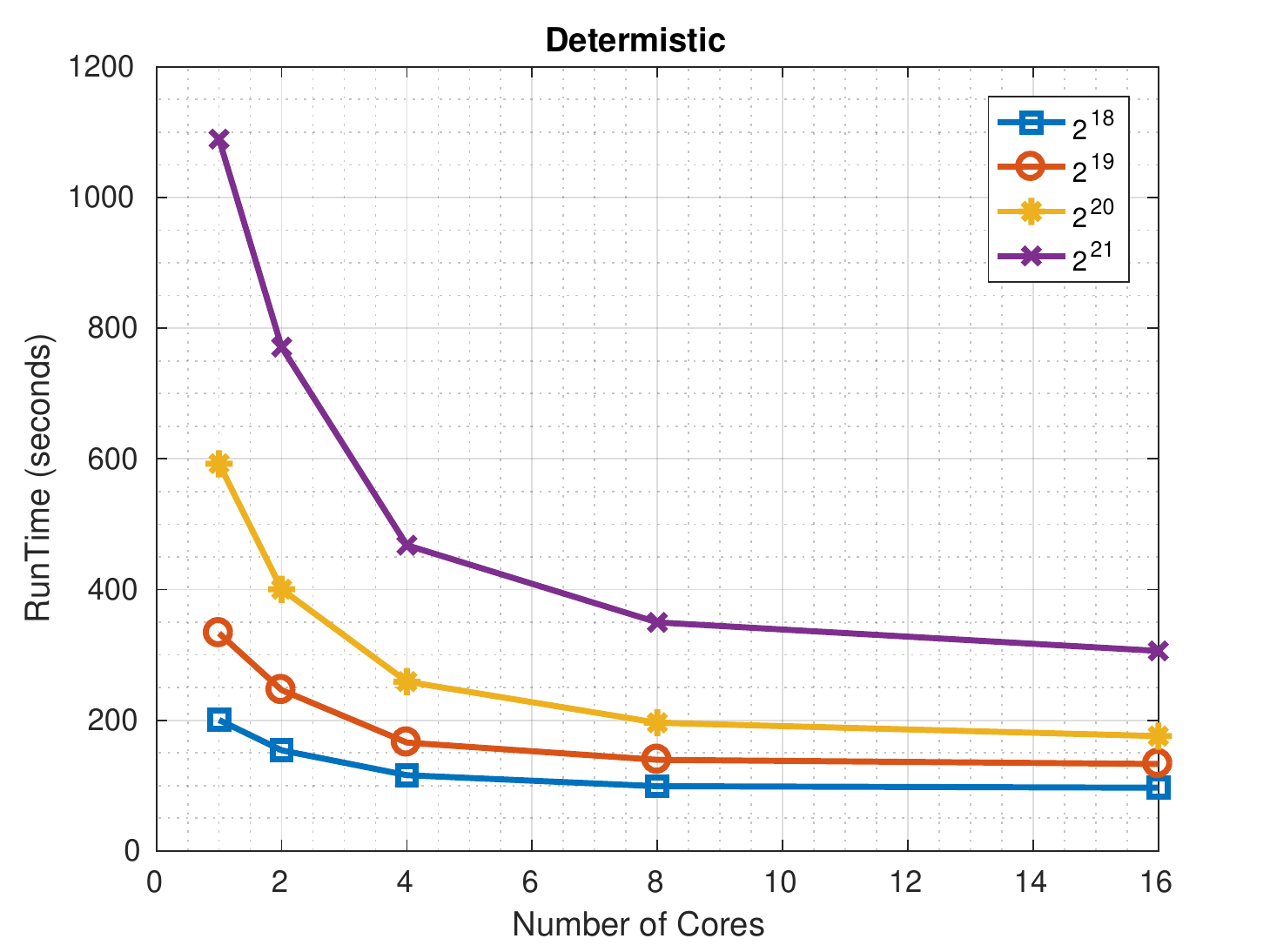} \\
    (b) Proposed approach\\
\hline
\end{tabular}
\caption{Worst-case performance of Redistribution: Platform 1.\label{fig:worst-case-runtime-p1}}
\end{center}
\end{figure}

\begin{figure}
\begin{center}
\begin{tabular}{|c|}
\hline
    \includegraphics[width=0.44\textwidth]{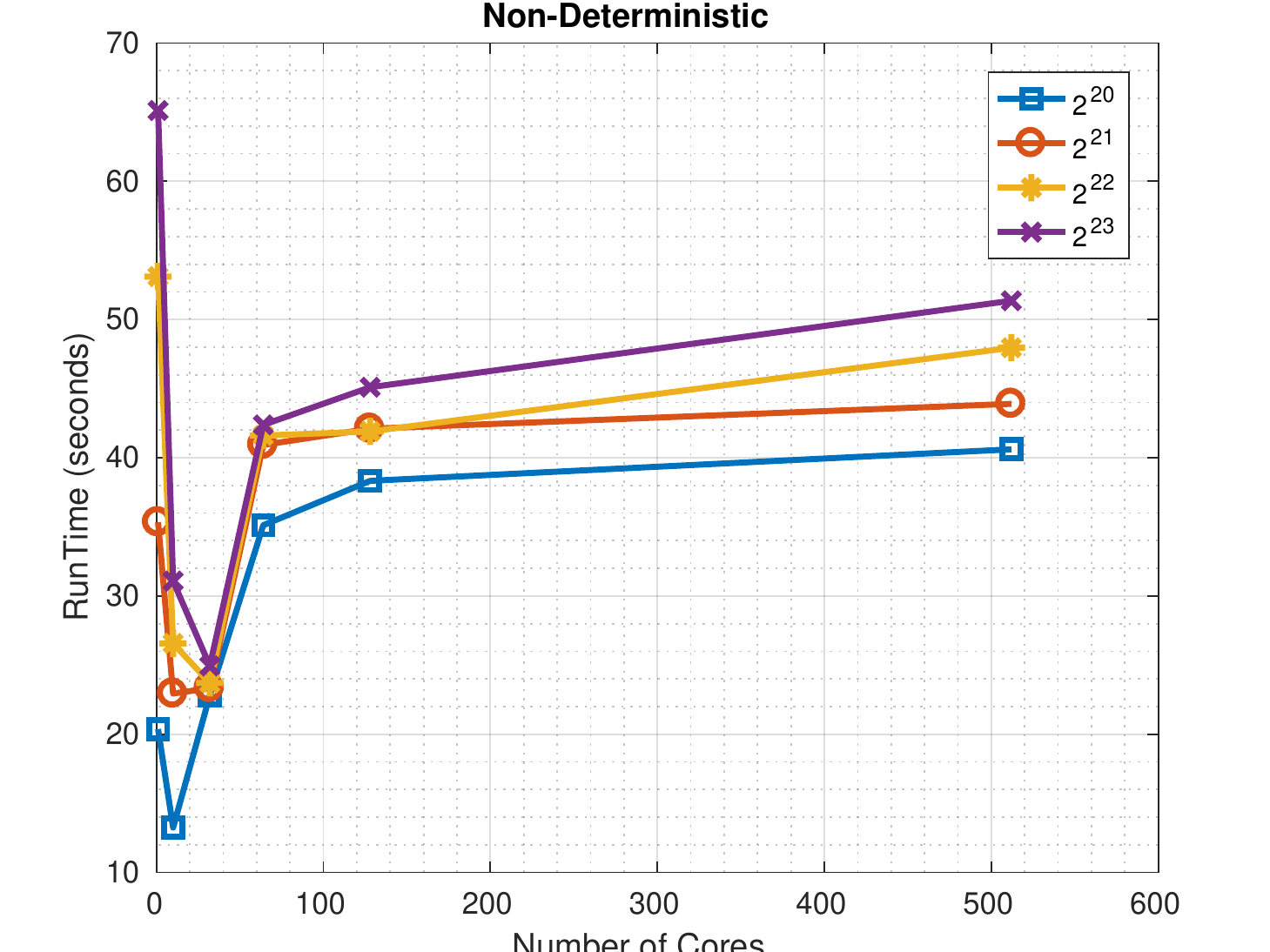} \\
    (a) Na\"{i}ve implementation\\
\hline
\end{tabular}
\quad
\begin{tabular}{|c|}
\hline
    \includegraphics[width=0.44\textwidth]{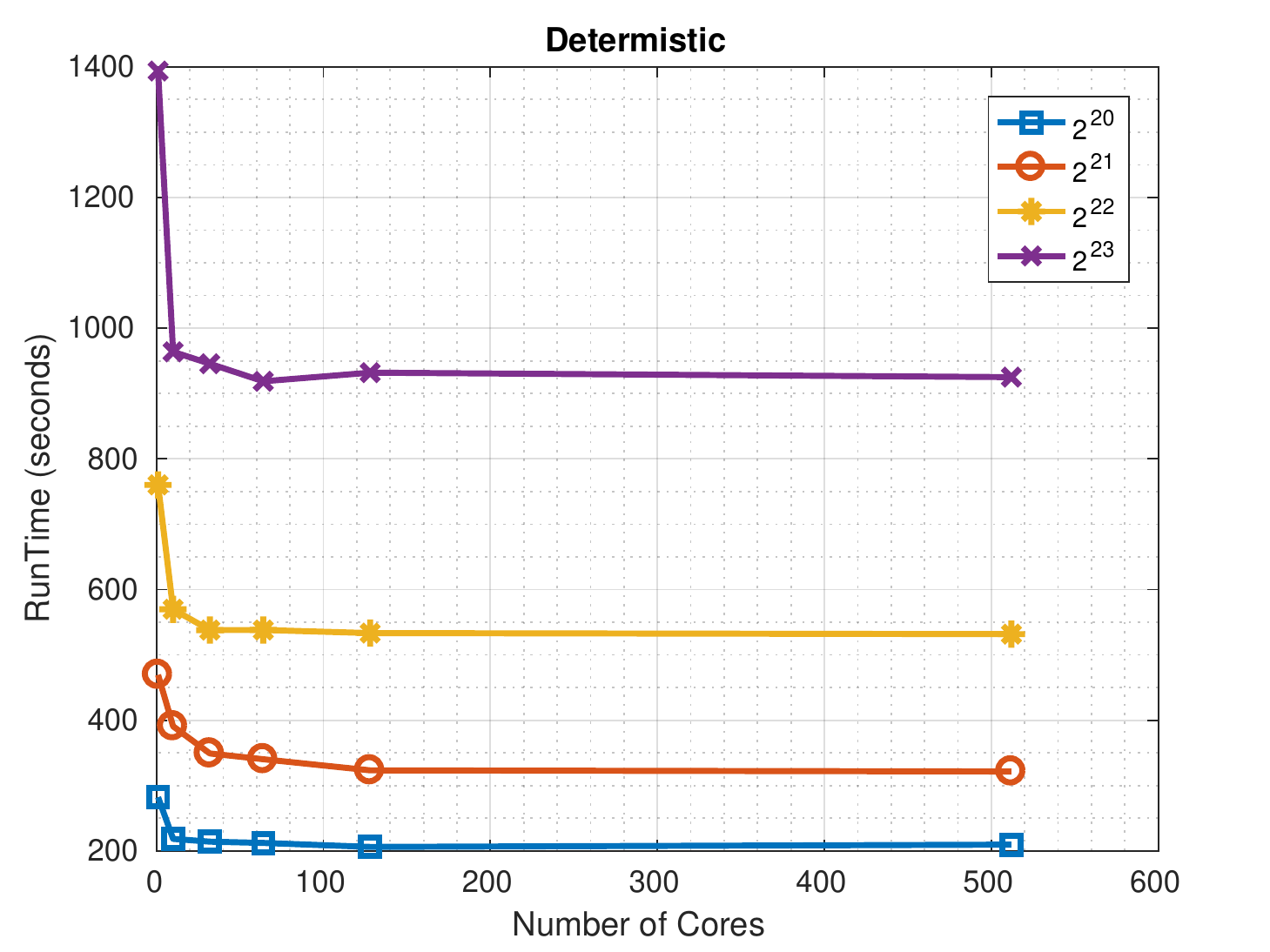} \\
    (b) Proposed approach\\
\hline
\end{tabular}
\caption{Worst-case performance of Redistribution: Platform 2.\label{fig:worst-case-runtime-p2}}
\end{center}
\end{figure}

We investigated the worst-case performance of such a na\"{i}ve parallel implementation of the redistribution component and compared with our proposed implementation (using a Spark implementation). The results are shown in figures~\ref{fig:worst-case-runtime-p1} and~\ref{fig:worst-case-runtime-p2} for the worst-case (where the new population of particles are all copies of a single member of the old population). It should be evident that as the number of cores increases, the runtime of the proposed (almost) never increases\footnote{In subsequent sections, we will investigate how and when the decrease in runtime occurs in more detail.}. In contrast, while the runtime of the na\"{i}ve implementation initially decreases as the number of cores is increased, it then increases (i.e., such that it is faster in absolute terms to use 8 not 16 cores with Platform 1 and such that it is faster to use less than 50 cores not 512 cores with Platform 2). The reason for the decrease is that the map-reduce framework can use the extra cores to more rapidly process the (many) zeros in the vector describing the number of copies. The reason for the subsequent increase in processing time is that the additional overhead of having multiple cores becomes increasingly significant if only one of the cores is doing the vast majority of the processing.

It should also be evident that the absolute runtime (on these platforms and with our current Spark implementation) of the deterministic and non-determinsitic variants differ significantly such that the na\"{i}ve implementation can be approximately 20 times faster (in the contexts of both platforms). This is disappointing and does motivate future work to refine our (initial) implementation. However, we perceive that there are applications where a slower but deterministic runtime is preferable to a faster but data-dependent runtime. In the contexts of such applications, particularly given the scope to improve the implementation, we perceive our algorithm (if not our current implementation) has utility.

\subsection{Overall Profile}\label{sec:profile}

\begin{figure}
    \centering
    \begin{adjustbox}{varwidth=\textwidth,fbox,center}
      \includegraphics[width=\textwidth]{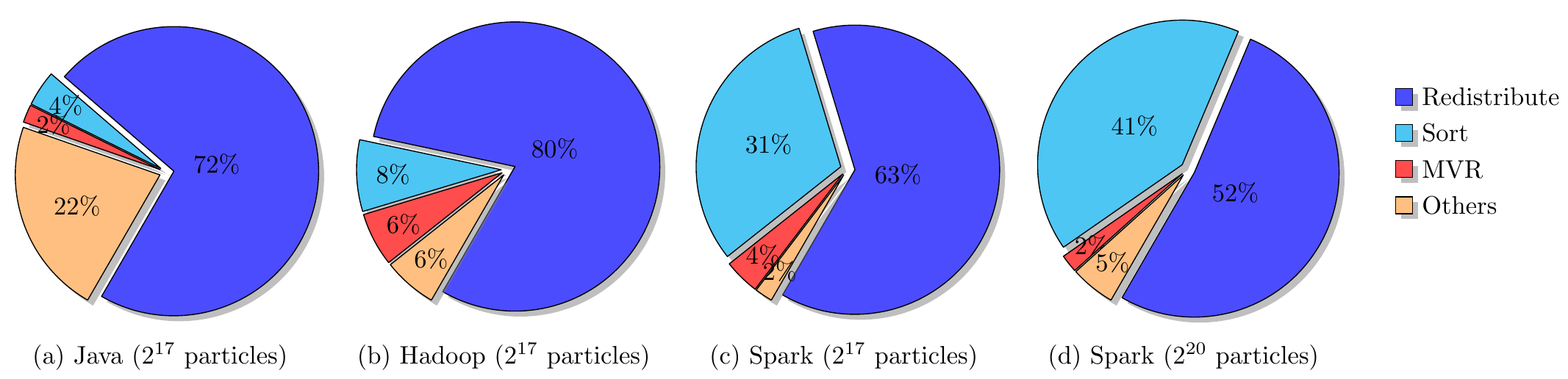}
    \end{adjustbox}
    \caption{Overall  runtime  profile  of  the  particle  filtering  algorithm  for the following implementations: (a)  Sequential; (b)
Hadoop; (c) Spark with $2^{17}$ particles; (d) Spark with $2^{20}$ particles.}
    \label{fig:overall-profile}
 \end{figure}

We now compare performance of three implementations of a particle filter: a sequential implementation (in Java and using quicksort in place of bitonic sort); an implementation in Hadoop; an implementation in Spark. All implementations involve a single core and Platform 1. Figure~\ref{fig:overall-profile} shows the proportion of the runtime that is associated with: redistribution; sort; Minimum Variance Resampling (MVR); the remaining components (e.g., sum, cumulative sum, diff, scaling).

As can be observed from figure~\ref{fig:overall-profile}, the majority of the time taken is devoted to the redistribution component. Note that, for the Spark implementation, a significant fraction of the remaining time is spent on the sorting component and the fraction of time devoted to redistribution and sorting increases as the number of particles is increased.



\subsection{Comparison of Hadoop and Spark}\label{sec:hadoopspark}

We next investigate how the choice of middleware impacts performance in the context of the
components of the algorithm and in the context of the entire particle filter algorithm. All implementations involve a single core and Platform 1.

\subsubsection{Sum and Cumulative Sum}\label{sec:results:sumcumsum}

\begin{figure}
\begin{tabular}{|c|}
\hline
    \includegraphics[width=0.44\textwidth]{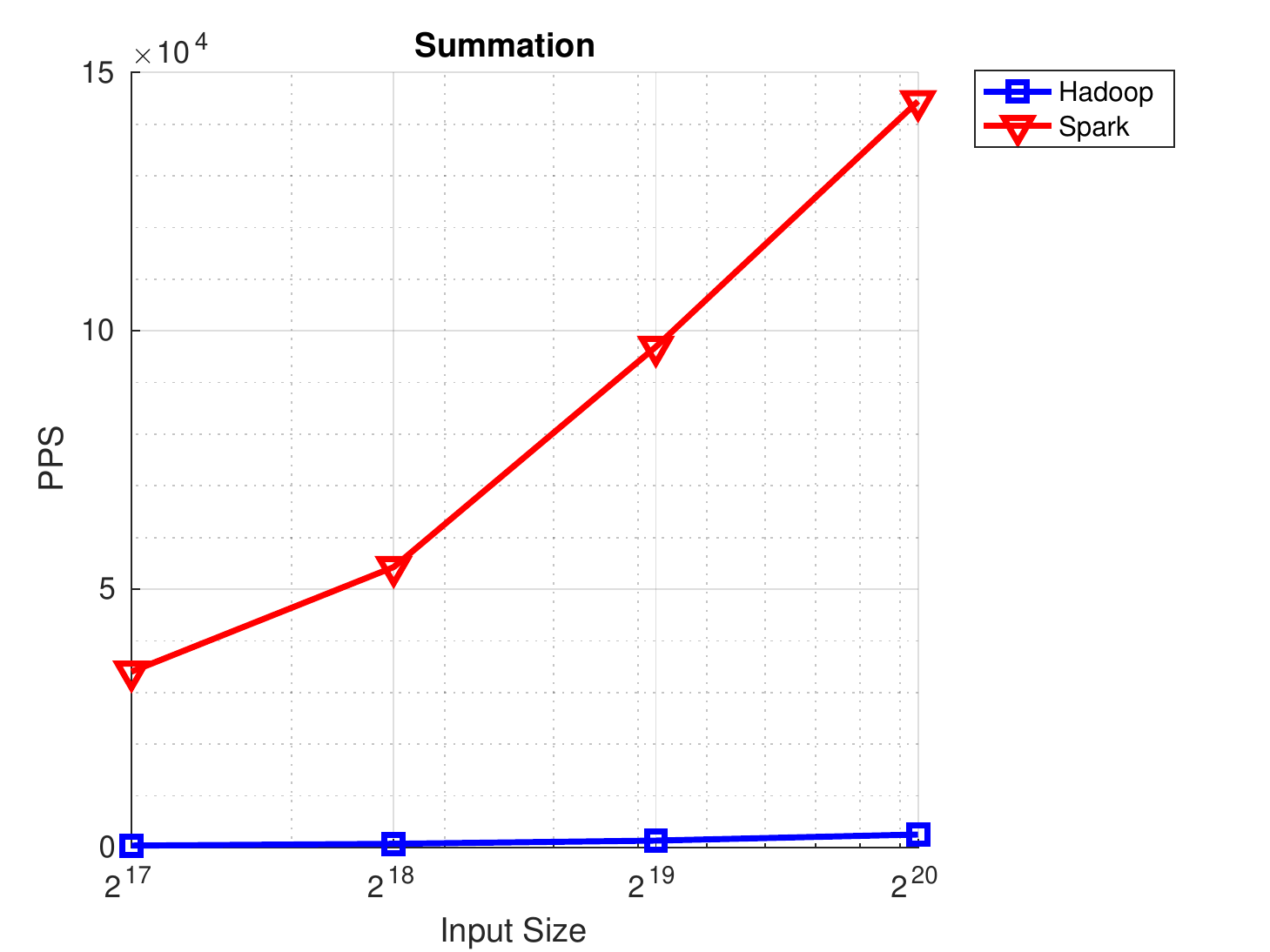} \\
    (a) Summation \\
\hline
\end{tabular}
\quad
\begin{tabular}{|c|}
\hline
    \includegraphics[width=0.44\textwidth]{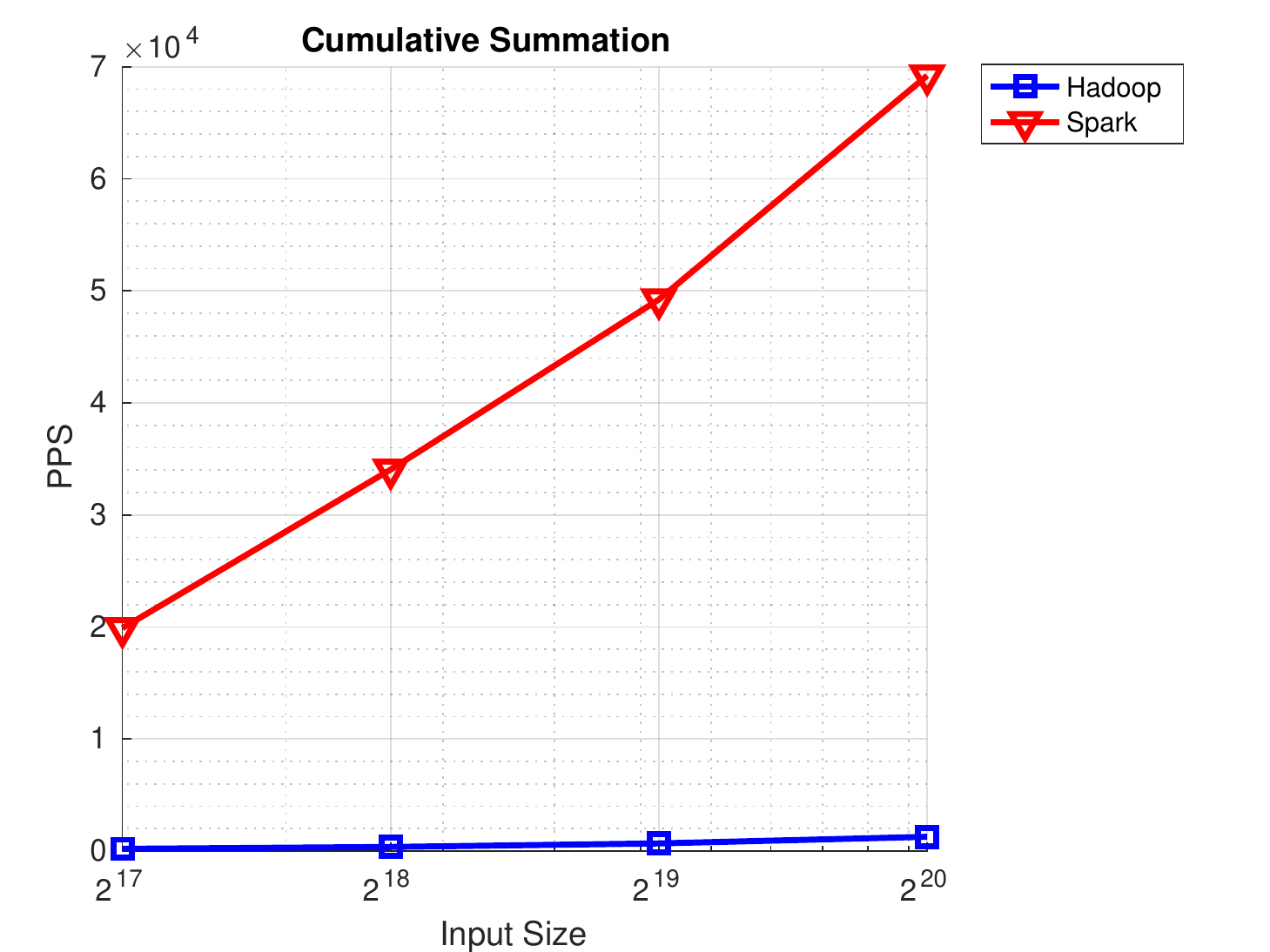} \\
    (b) Cummulative Summation \\
\hline
\end{tabular}
\caption{Summation and Cumulative Summation on Spark and Hadoop.}
\label{fig:pps-components-sums}
\end{figure}

Figure~\ref{fig:pps-components-sums} shows the comparative
performance of the sum and cumulative sum components in these two key
frameworks.

With respect to the number of particles processed per second (PPS), the performance using Spark is far superior to that achieved using Hadoop. This stems from the issues discussed in section~\ref{sec:bigdata}: Spark uses RDDs to makes use of memory (and lazy evaluation) whereas Hadoop only uses the file system (HDFS) to transfer data from the output of one operation to the input of the next.

It is apparent in both frameworks (and particularly apparent in the context of Spark) that, as the number of particles increases, the number of particle processed per second also increases. This is because with more particles, the overheads associated with setting up (and tearing down) the mappers and reducers are increasingly offset by the parallel operations that make use of the mappers and reducers. The limited extent to which this effect is observed in the context of Hadoop highlights that the overheads associated with opening files in HDFS are significant.

Since, as explained in section~\ref{sec:components}, calculating a summation involves one adder tree and cumulative sum involves two such trees, we should expect the number of particles per second for the cumulative sum to be approximately half that for the summation. A comparison of the two graphs in figure~\ref{fig:pps-components-sums} makes clear that this is indeed (approximately) the case for both frameworks and for all input sizes.

\subsubsection{Bitonic Sort and Minimum Variance Resampling}

\begin{figure}
\begin{tabular}{|c|}
\hline
    \includegraphics[width=0.43\textwidth]{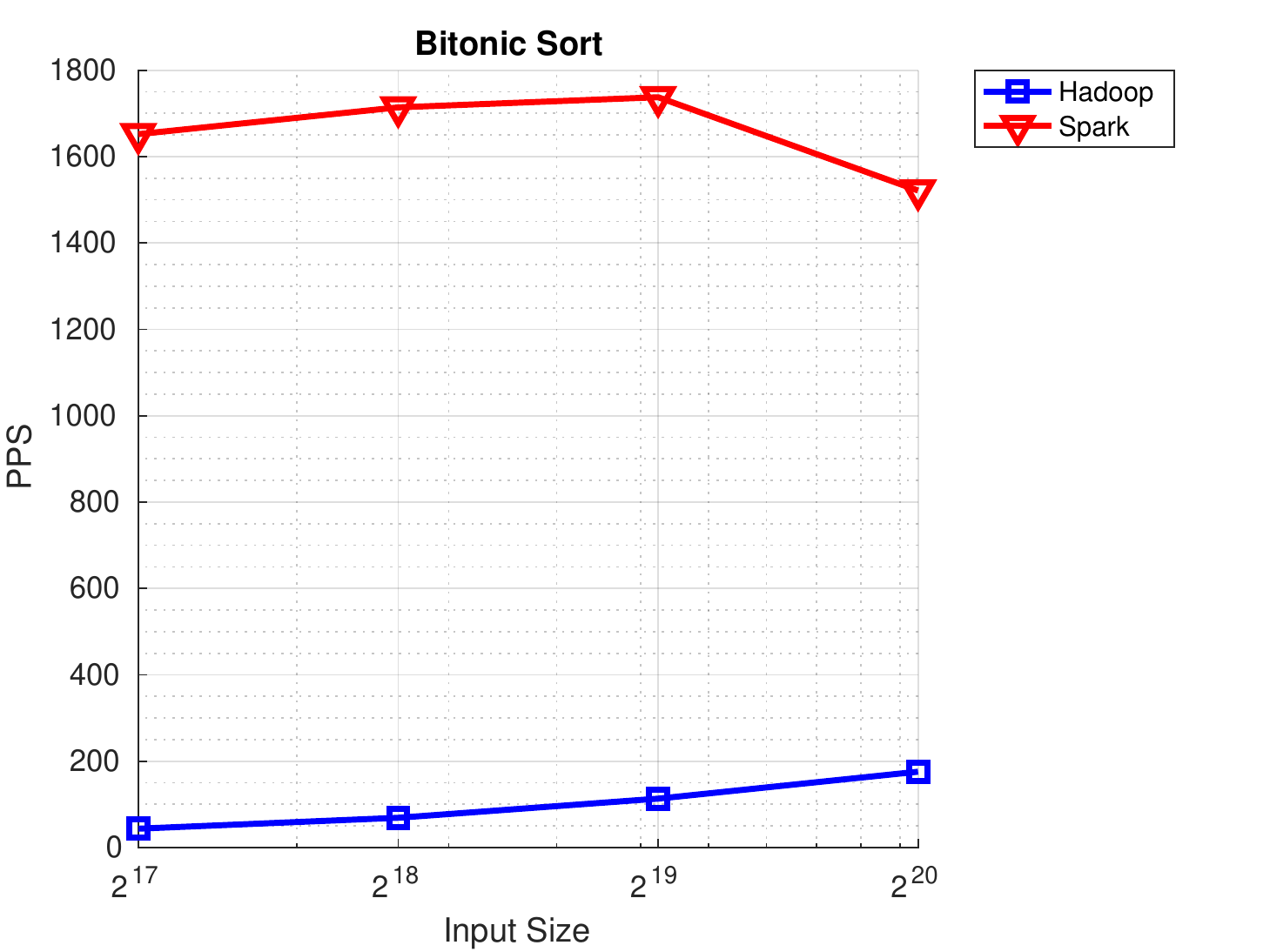} \\
    (a) Bitonic Sort \\
\hline
\end{tabular}
\quad
\begin{tabular}{|c|}
\hline
    \includegraphics[width=0.43\textwidth]{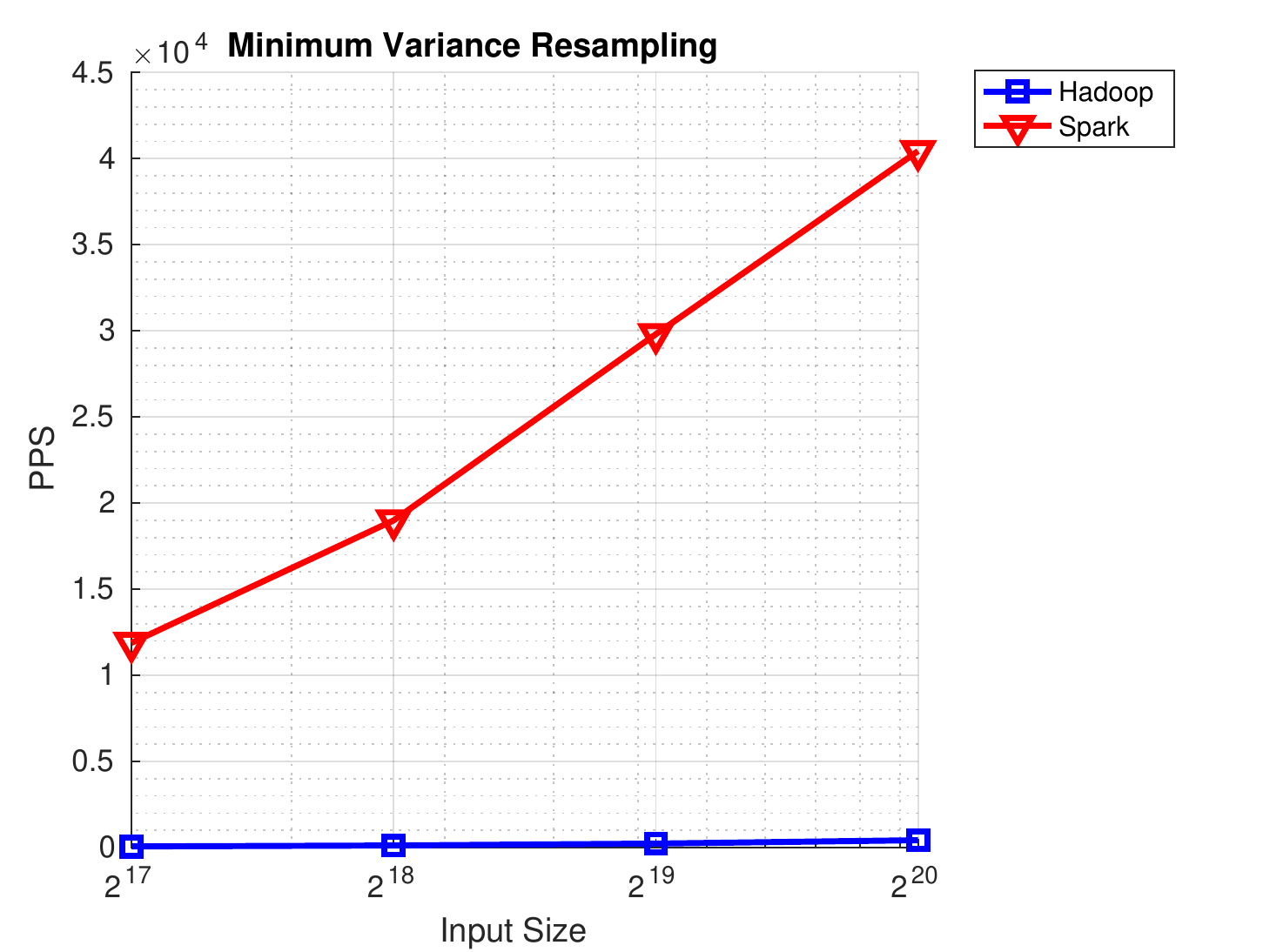} \\
    (b) Minimum Variance Resampling \\
\hline
\end{tabular}
\caption{Bitonic Sort and Minimum Variance Resampling on Spark and Hadoop.}
\label{fig:pps-components-sortmvr}
\end{figure}

Figure~\ref{fig:pps-components-sortmvr} shows the performance for two independent components, {\em bitonic sort} and {\em minimum variance resampling}. The performance of {\em minimum variance resampling} is relatively close to the performance of the cumulative sum~(see Figure \ref{fig:pps-components-sums}). This is expected since, as explained in section~\ref{sec:components}, {\em minimum variance resampling} includes a cumulative sum.

Once again and for the same reasons as discussed in section~\ref{sec:results:sumcumsum}, we notice the same difference in performance between the Spark and Hadoop implementations. As one might expect and as before, for {\em minimum variance resampling}, the number of particles per second increases with the number of particles. However, it is noteworthy that, for bitonic sort with Spark, the number of particles per second decreases for large numbers of particles. On investigating this in some detail, we observed that the \emph{lineages} used to facilitate the lazy evaluation in Spark\footnote{Since Hadoop does not attempt lazy evaluation or use such lineages for another purpose, the same phenomenon is not observed in the context of Hadoop.} become very large with large numbers of particles. This appears to cause Spark to become less efficient when the number of particles becomes large.


\subsubsection{Redistribution and Overall Performance}

\begin{figure}
\begin{tabular}{|c|}
\hline
    \includegraphics[width=0.43\textwidth]{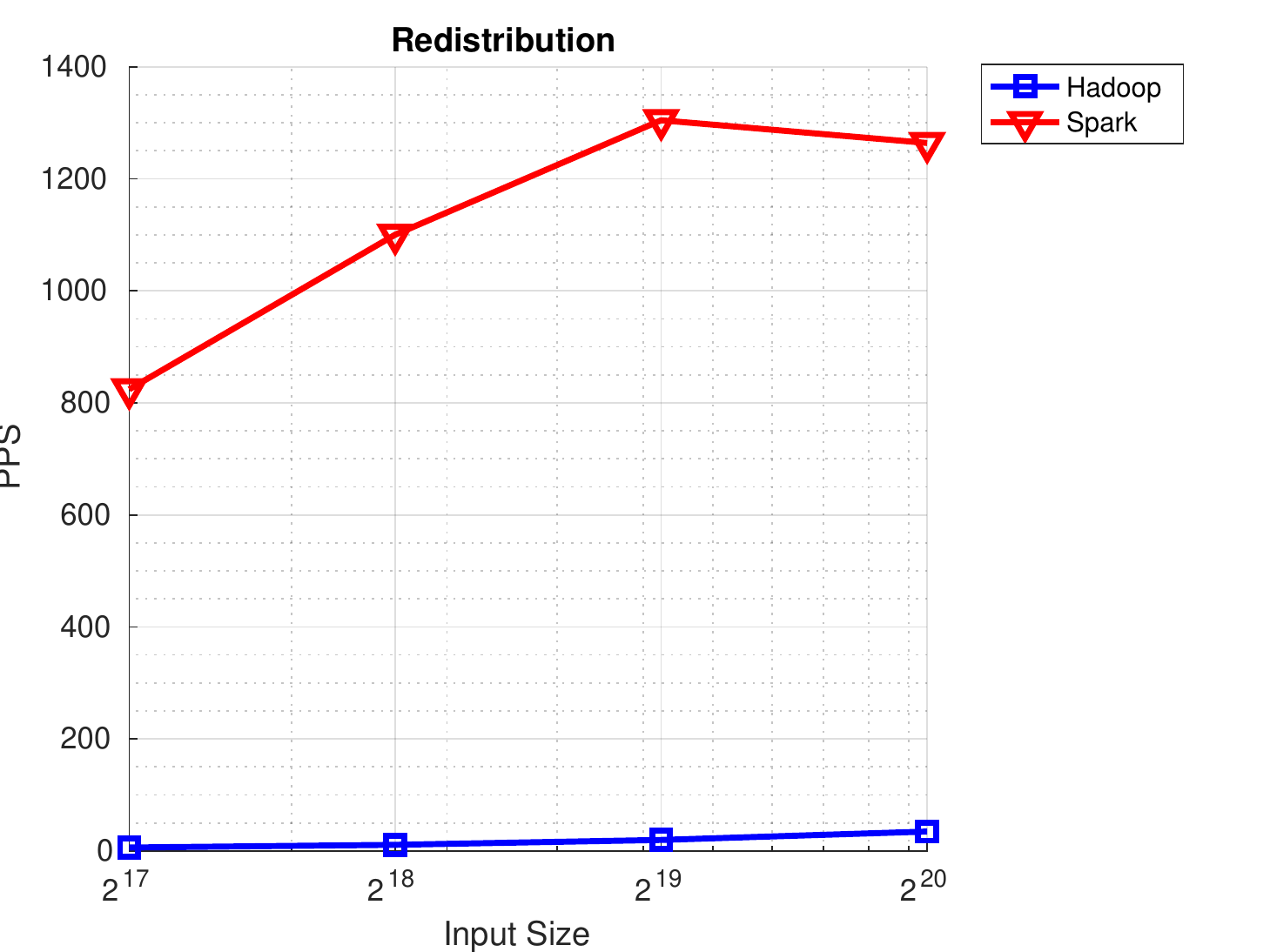}  \\
    (a) Redistribution \\
\hline
\end{tabular}
\quad
\begin{tabular}{|c|}
\hline
    \includegraphics[width=0.43\textwidth]{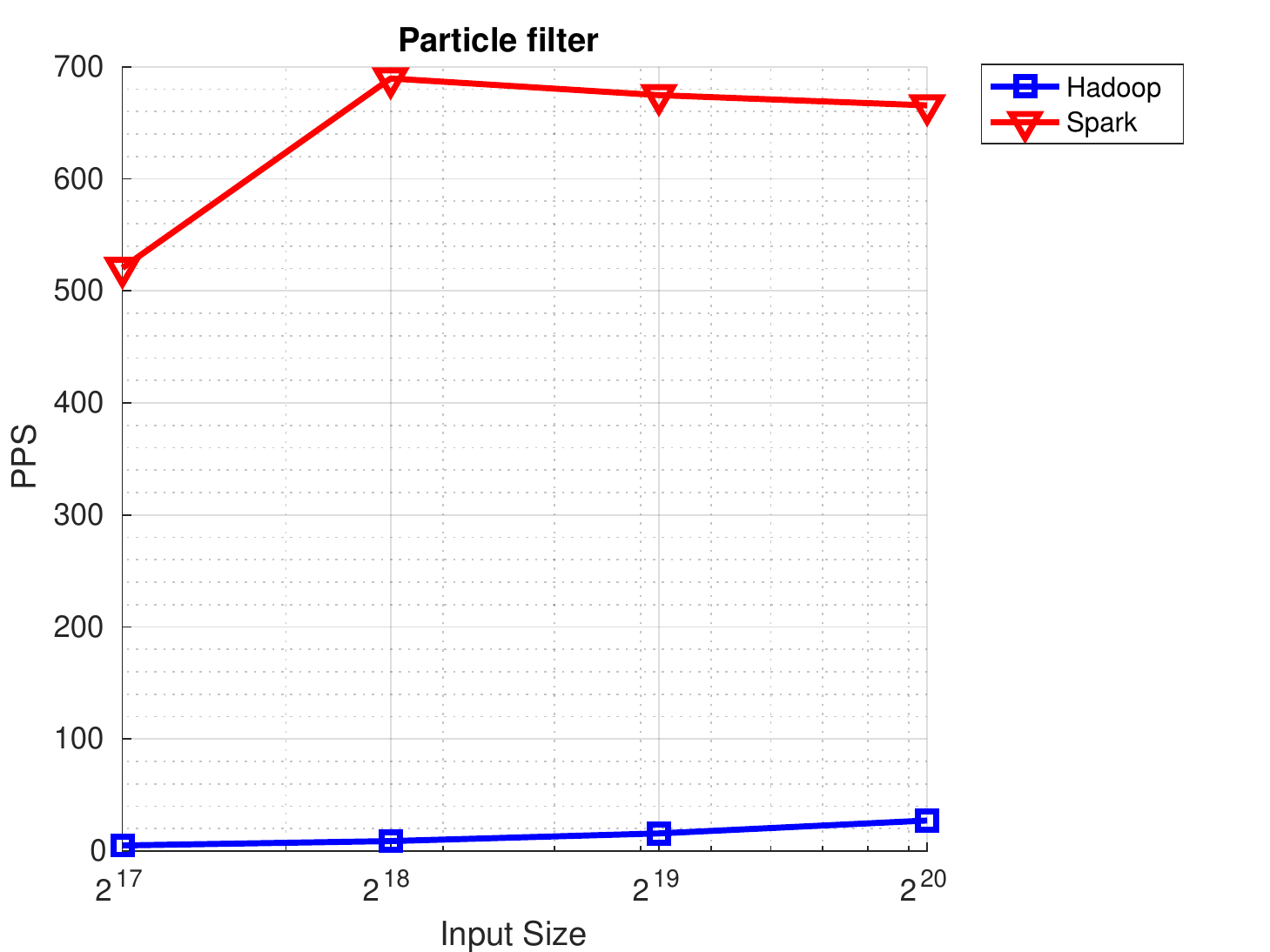} \\
    (b) Overall Particle Filtering \\
\hline
\end{tabular}
\caption{Redistribution and the Overall Particle Filtering on Spark and Hadoop.}
\label{fig:pps-components-pf-redistribute}
\end{figure}

Finally, figure~\ref{fig:pps-components-pf-redistribute} shows the comparative performance of the redistribution algorithm (as described in algorithm~\ref{alg:shift}) and the overall particle filtering algorithm.

Once again, we notice the same differences between Hadoop and Spark. In the context of the overall particle filter and for the largest number of particles considered, these differences are manifest in Spark, relative to Hadoop, offering a considerable speedup (approximately 25-fold\footnote{In the particle filter the resampling is executed in every iteration. Thus the aforementioned figures correspond to a worst-case speedup.}).


The overall performance of the particle filtering algorithm, when implemented in Spark, decreases for large numbers of particles. Again, on investigation, this appears to be caused by large lineages associated with the large number of particles. Finally, we note that the bitonic sort and redistribution components appear to be limiting the number of particles per second that can be processed by the overall particle filtering algorithm.


\subsection{Impact of Using Multiple Cores}
\label{sec:evlaution:multicore}

We now focus on the Spark implementation (with Platform 1) and compare the performance of the two variants of the redistribution component in isolation and in the context of the overall performance of a particle filter. More specifically, we investigate how performance scales with the number of cores and the number of particles.

\subsubsection{Redistribution Component in Isolation}

\begin{figure}
\begin{tabular}{|c|}
\hline
    \includegraphics[width=0.43\textwidth]{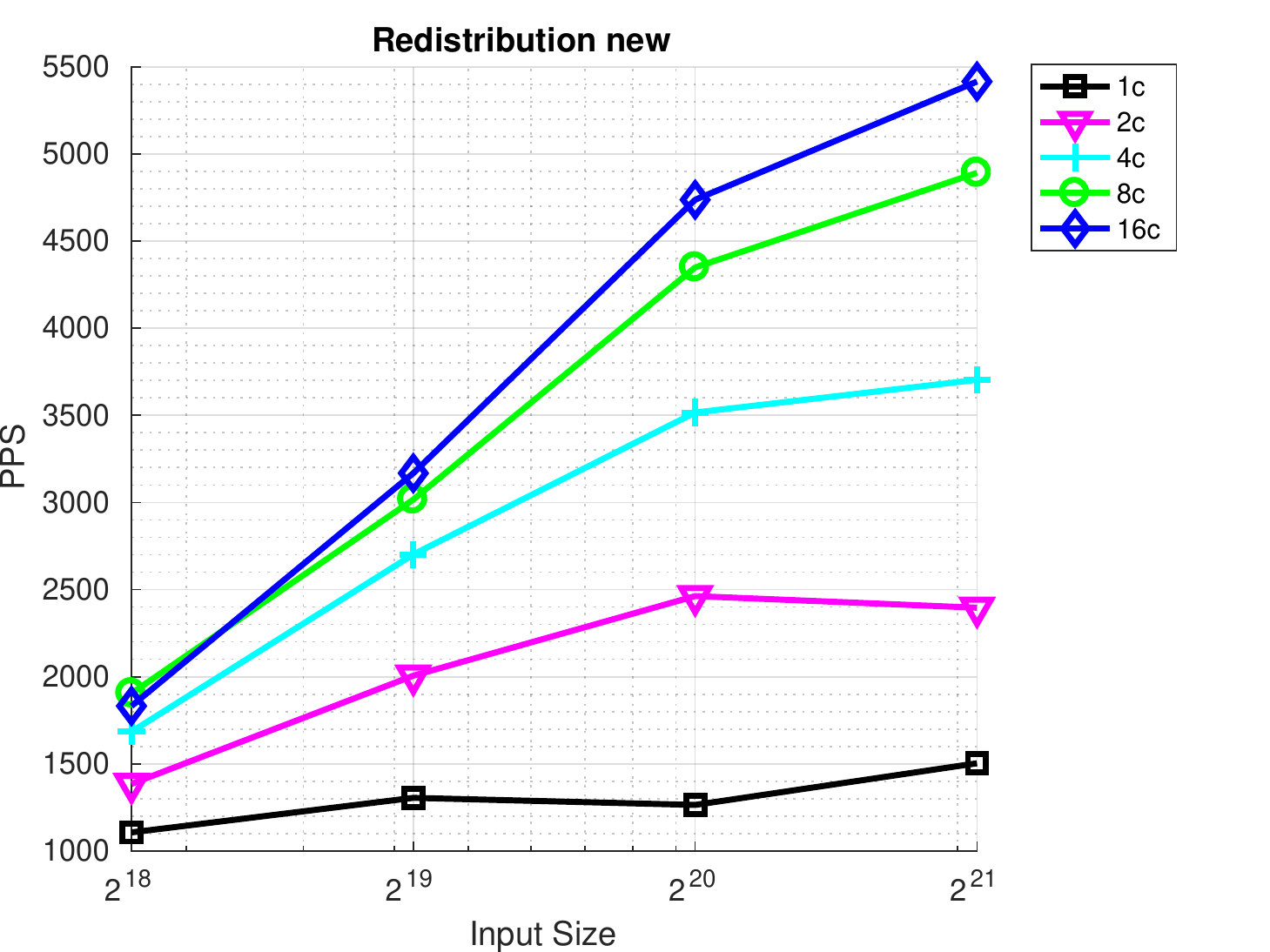}   \\
    (a) Redistribution $\bigo{(log_2N)^2}$ \\
\hline
\end{tabular}
\quad
\begin{tabular}{|c|}
\hline
    \includegraphics[width=0.43\textwidth]{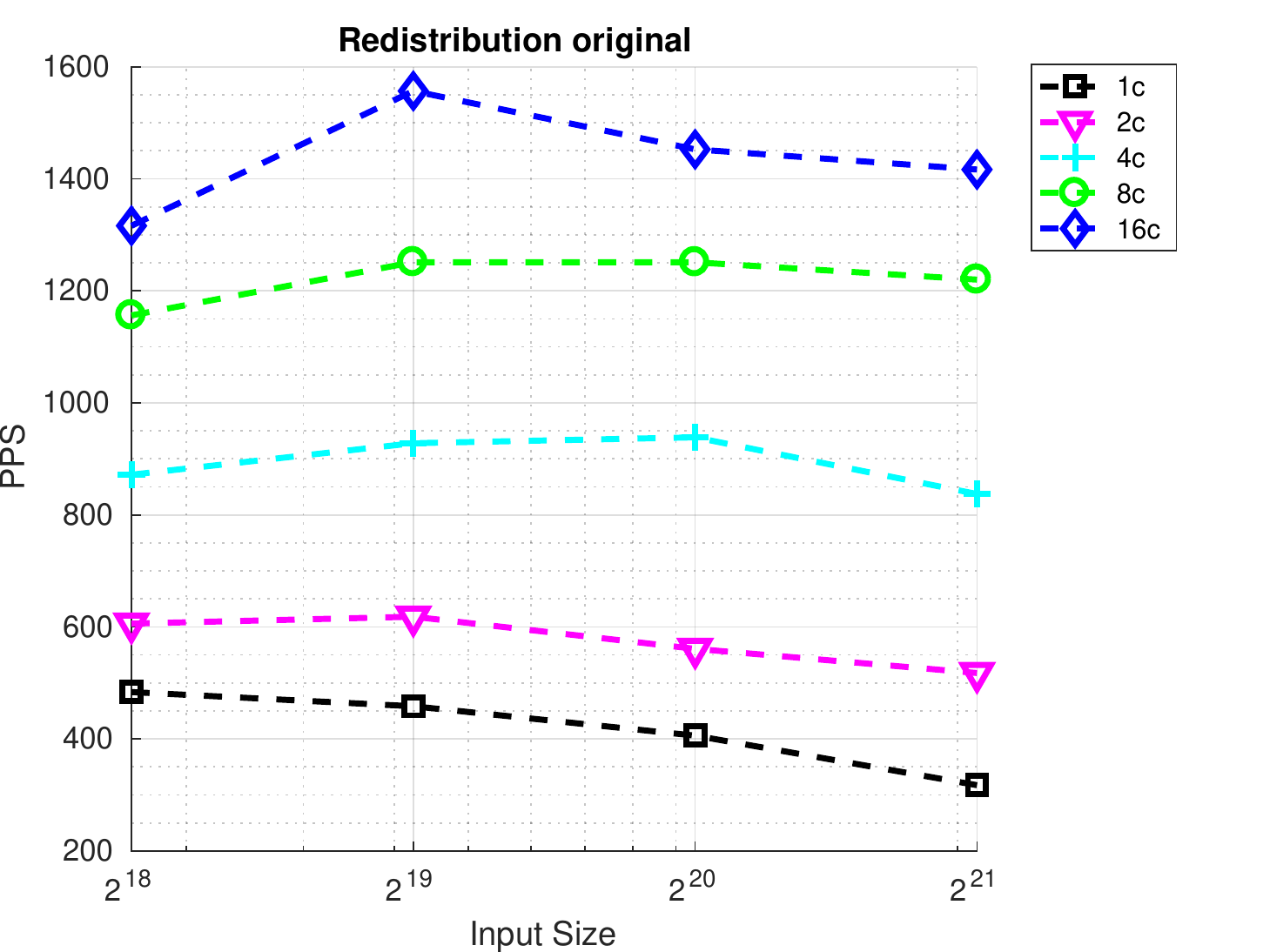} \\
    (b) Redistribution $\bigo{(log_2N)^3}$ \\
\hline
\end{tabular}
\caption{Performance of the two variants of the Redistribution Component (using Spark).}
\label{fig:pps-redistribution-spark-old-new}
\end{figure}

Figure~\ref{fig:pps-redistribution-spark-old-new} compares the
performance of the two versions of the redistribution
component as a function of the number of particles and number
of cores.

On a core-to-core basis, the $\bigo{\left(\log_2 N\right)^2}$ redistribution component outperforms the $\bigo{\left(\log_2 N\right)^3}$ component across all numbers of particles by a margin  of up to a factor of approximately 4 (for 16 cores). 

For all numbers of particles, increasing the number of cores improves performance for both variants of the redistribution component. However, in the context of both variants, the improvement in performance when considered as a ratio is less than the ratio of the number of cores.

In the context of the $\bigo{\left(\log_2 N\right)^3}$ variant, increasing the number of particles for a fixed number of cores can significantly reduce the number of particles processed per second. This is not the case for the $\bigo{\left(\log_2 N\right)^2}$ variant.

For the $\bigo{\left(\log_2 N\right)^2}$  variant, increasing the number of particles while keeping the number of cores constant improves the number of particle processed
per  second.  However,  in  the  context  of  the $\bigo{\left(\log_2 N\right)^3}$  variant,  increasing the number of particles for a  fixed number of cores can reduce the number of particles processed per second.

\subsubsection{Resulting Overall Particle Filter Performance}

\begin{figure}
\begin{tabular}{|c|}
\hline
    \includegraphics[width=0.43\textwidth]{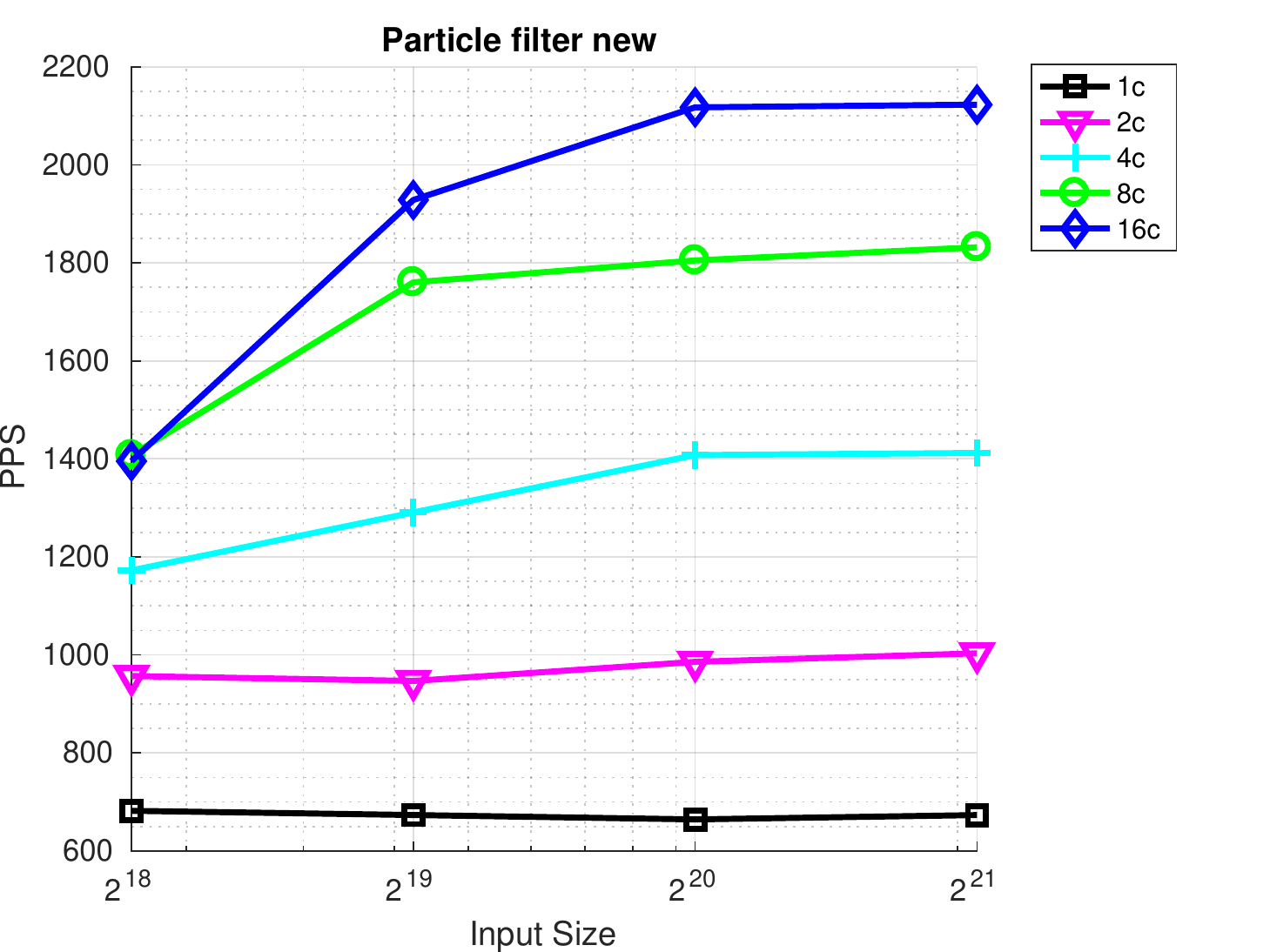} \\
    (a) Particle filter $\bigo{(log_2N)^2}$ \\
\hline
\end{tabular}
\quad
\begin{tabular}{|c|}
\hline
    \includegraphics[width=0.43\textwidth]{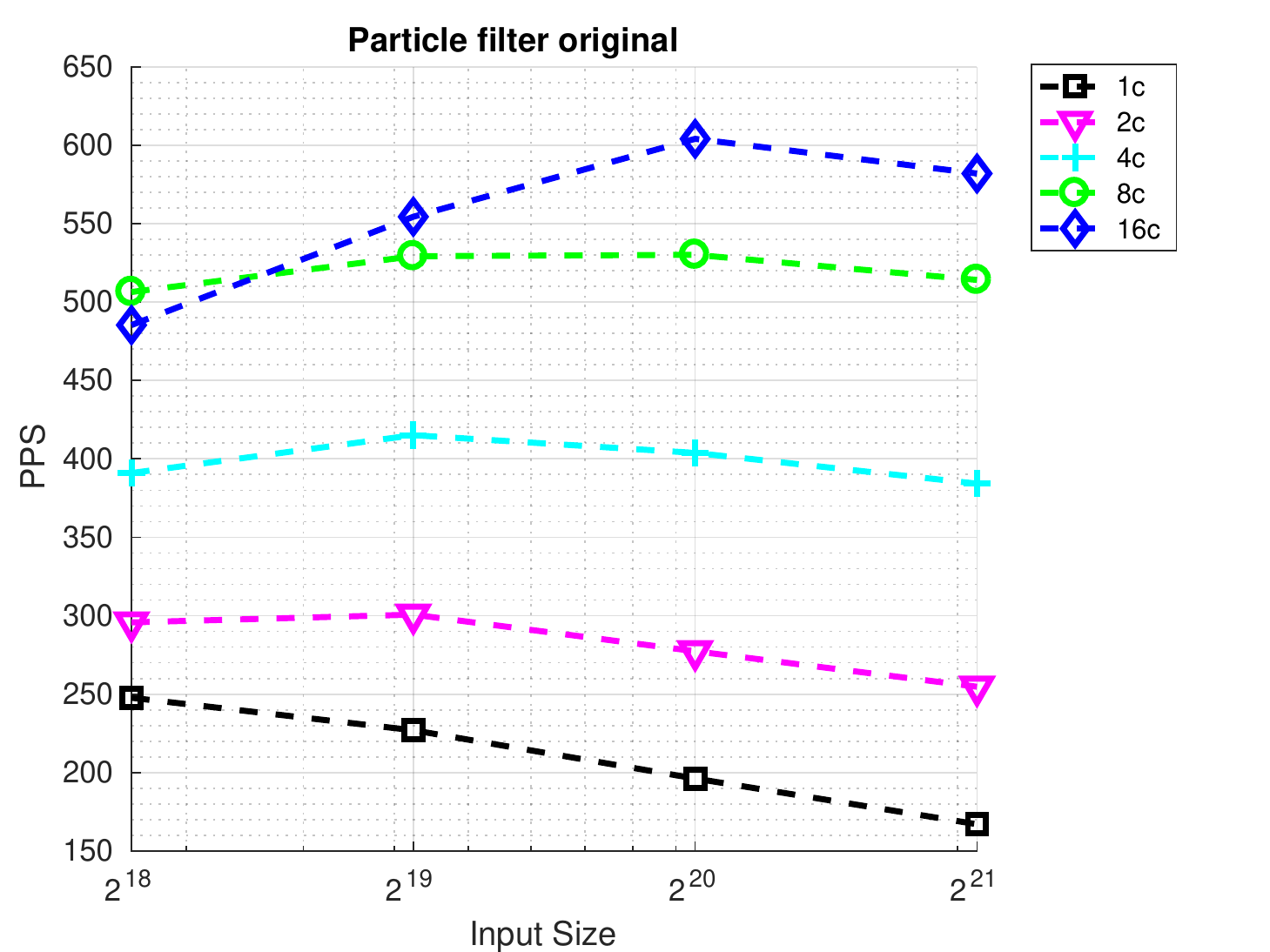} \\
    (b) Particle filter $\bigo{(log_2N)^3}$ \\
\hline
\end{tabular}
\caption{Performance of the overall particle filter using the two variants of the redistribution component.}
\label{fig:pps-pf-spark-old-new}
\end{figure}

Figure~\ref{fig:pps-pf-spark-old-new} compares the performance of the original particle filtering algorithm when using the two variants of the redistribution component.

The comparative performance that was observed in the context of the redistribution component in isolation is also evident when comparing the performance of the overall particle filter. Indeed, the use of the $\bigo{(log_2N)^2}$ variant of redistribution results in (approximately) a fourfold increase in the number of particles processed per second. The trends that were observed in the context of the redistribution component in isolation are also apparent in the context of the overall particle filter.

\subsection{Speedup and Scalability Analysis}\label{sec:speedup}

We now focus on the speedup that the $\bigo{(log_2N)^2}$ variant of the redistribution component offers relative to the $\bigo{(log_2N)^3}$ variant and the scalability of the $\bigo{(log_2N)^2}$ variant, i.e., the extent to which using more cores improves performance.

We quantify speedup as the ratio of the number of particles per second for a fixed number of particles and number of cores. We quantify scalability, in the context of a fixed number of particles\footnote{Since the problem size remains fixed, we are actually quantifying \emph{strong scaling}~\cite{Hill:scalability:1990}.}, as the ratio of the number of particles per second with $N$ cores relative to the number of particles per second with a single core.

We compare performance in the context of both platforms for various different numbers of particles.

\subsubsection{Redistribution Component in Isolation}

\begin{figure}
\begin{tabular}{|c|}
\hline
    \includegraphics[width=0.43\textwidth]{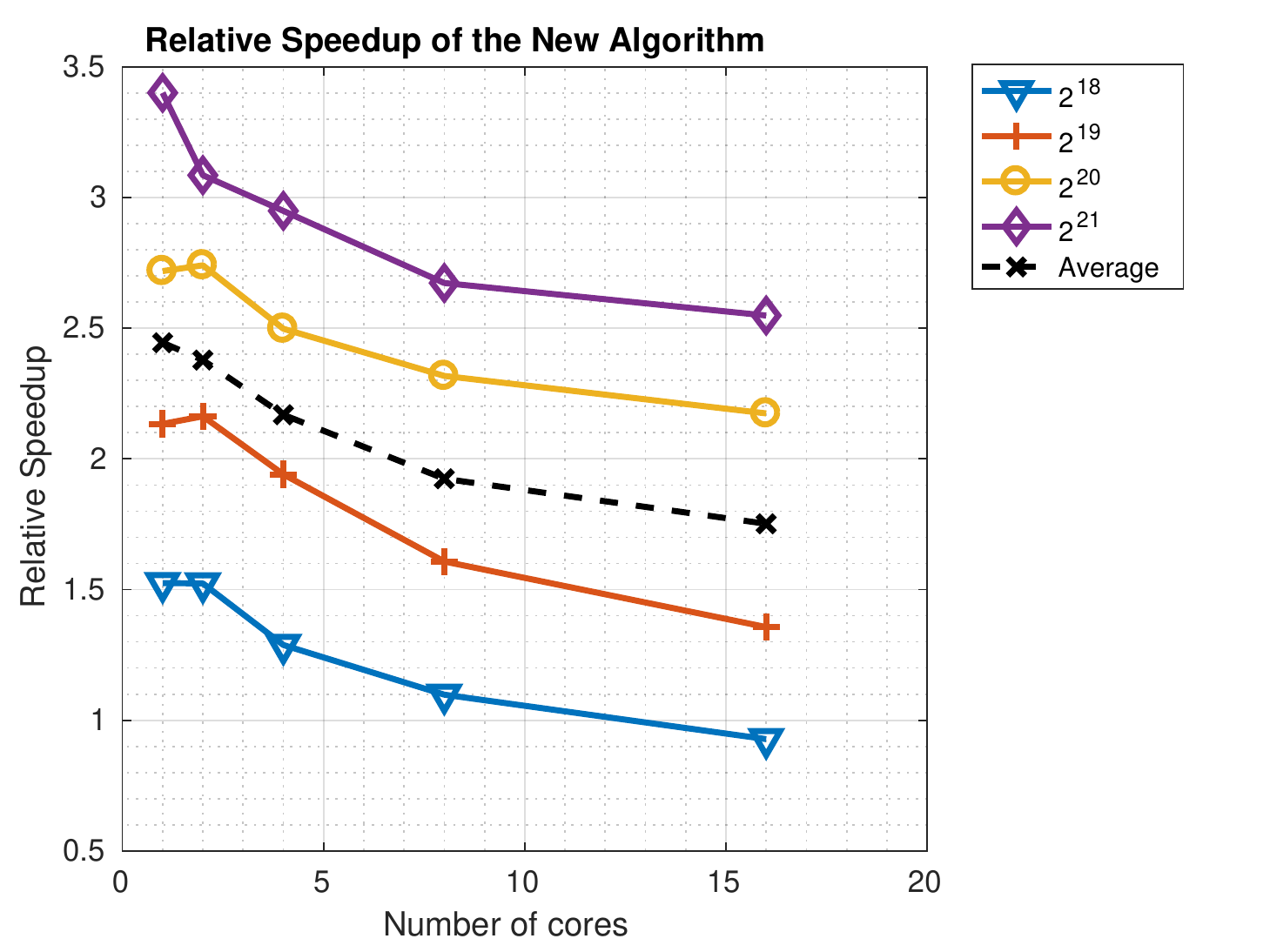}   \\
    (a) Relative Speedup of $\bigo{(log_2N)^2}$ variant\\
\hline
\end{tabular}
\quad
\begin{tabular}{|c|}
\hline
    \includegraphics[width=0.43\textwidth]{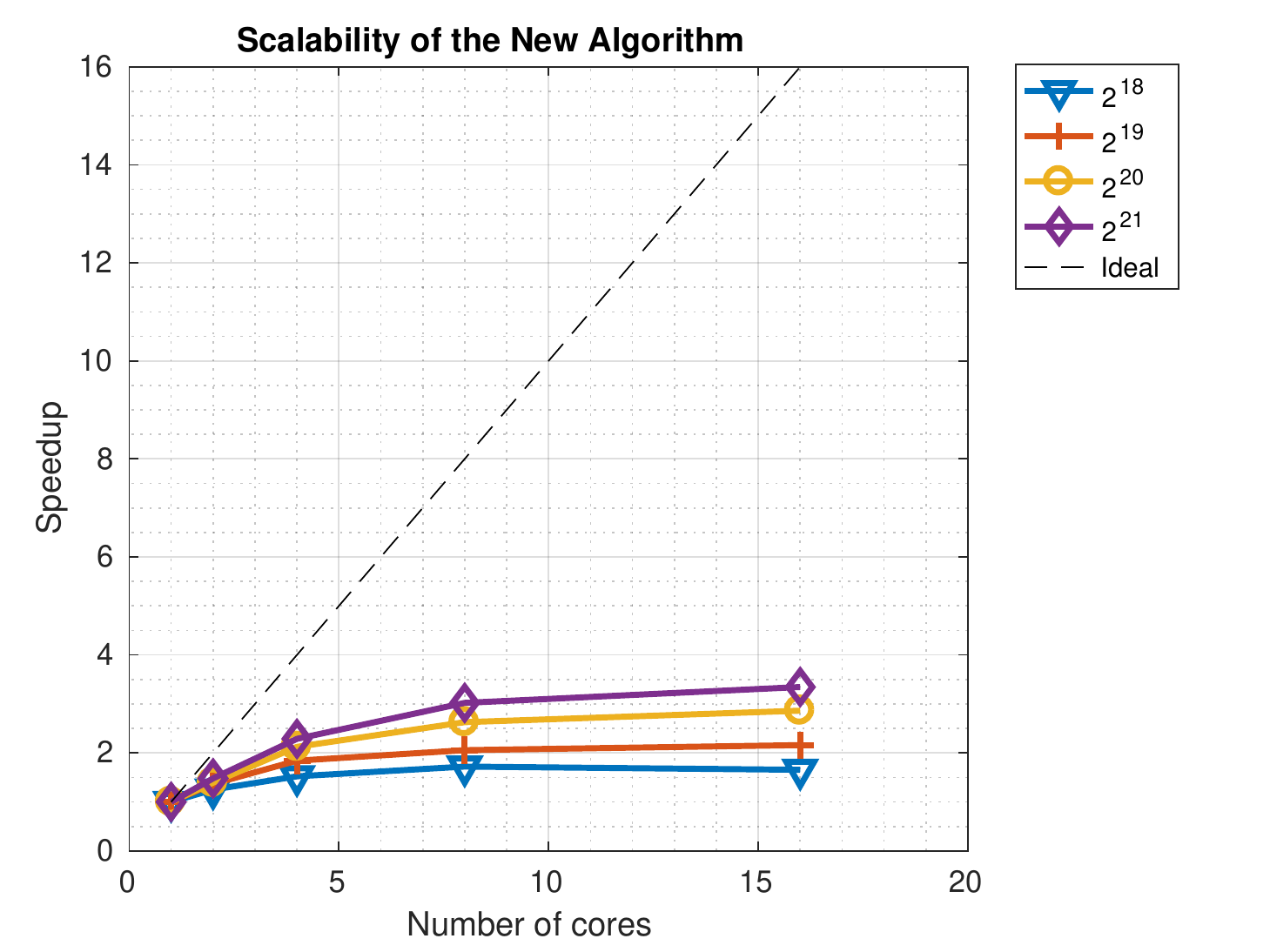} \\
    (b) Scalability of the $\bigo{(log_2N)^2}$ variant\\
\hline
\end{tabular}
\caption{Relative Speedup and Scalability of the $\bigo{(log_2N)^2}$ variant of the Redistribution component on Platform~1.}
\label{fig:redistribute-spark-speedup-scalability}
\end{figure}

\begin{figure}
\begin{tabular}{|c|}
\hline
    \includegraphics[width=0.43\textwidth]{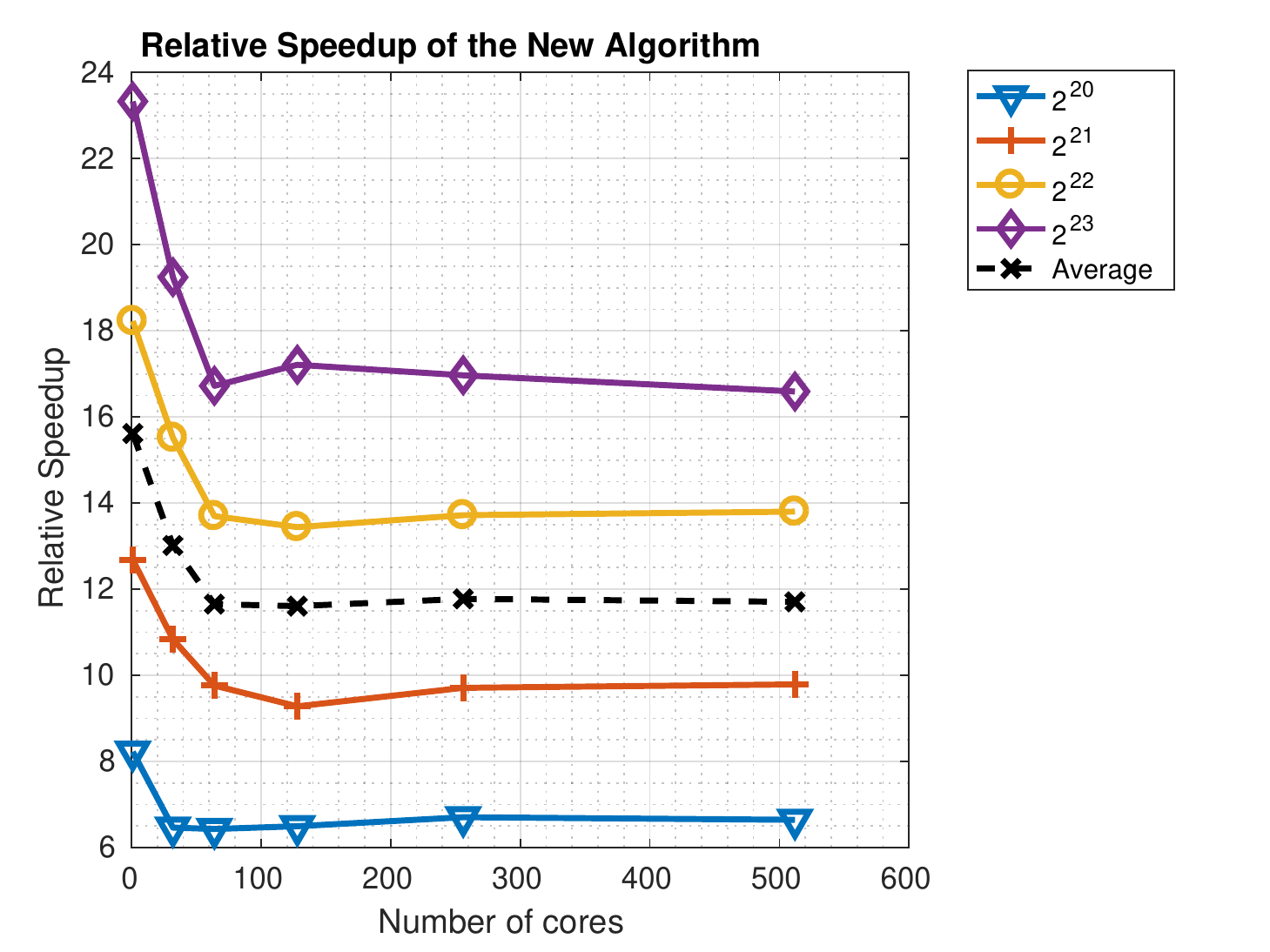}   \\
    (a) Relative Speedup of $\bigo{(log_2N)^2}$ variant\\
\hline
\end{tabular}
\quad
\begin{tabular}{|c|}
\hline
    \includegraphics[width=0.43\textwidth]{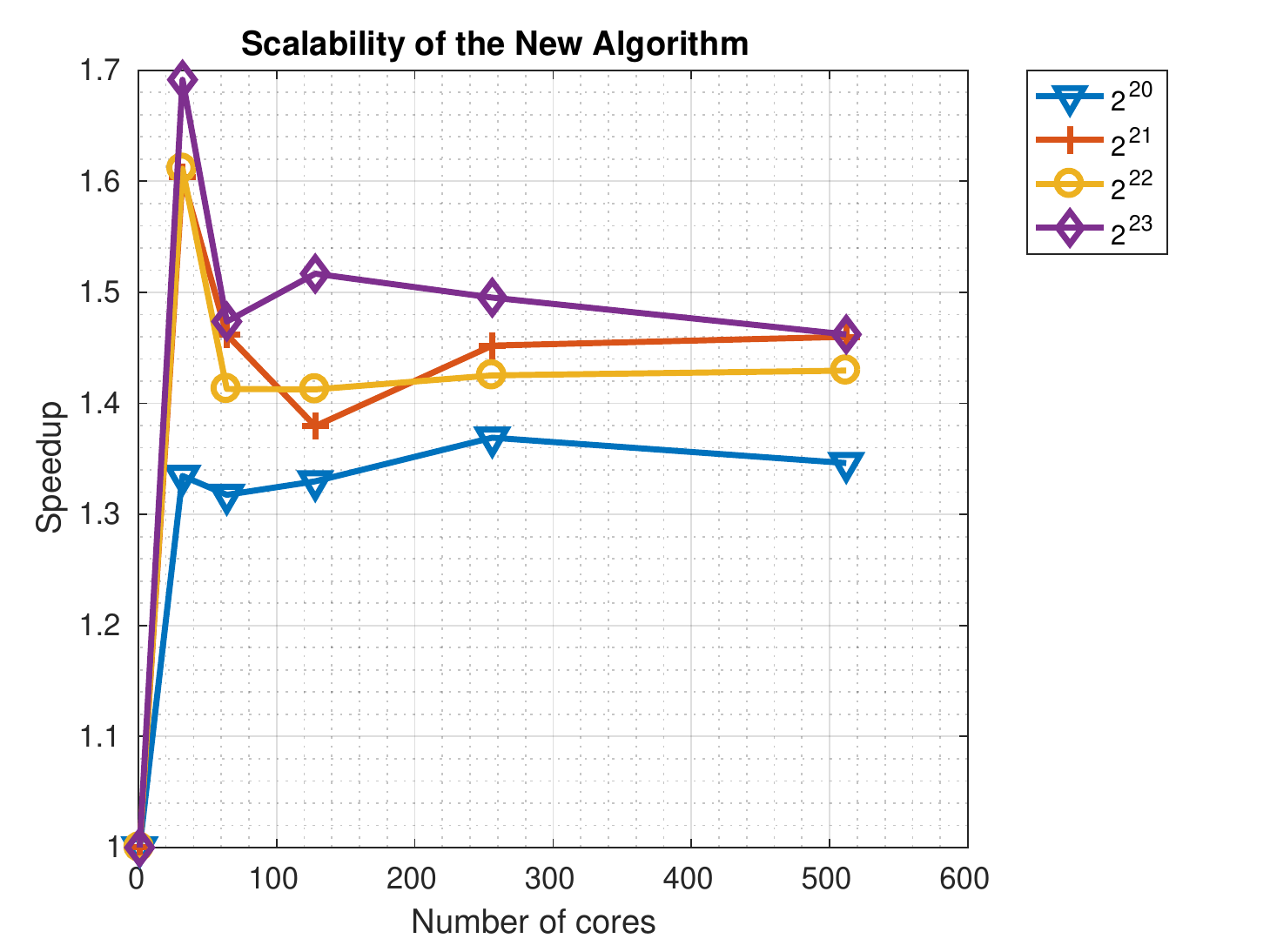} \\
    (b) Scalability of the $\bigo{(log_2N)^2}$ variant\\
\hline
\end{tabular}
\caption{Relative Speedup and Scalability of the $\bigo{(log_2N)^2}$ variant of the Redistribution component on Platform~2.}
    \label{fig:redistribute-spark-speedup-scalability2}
\end{figure}

Figures~\ref{fig:redistribute-spark-speedup-scalability} and~\ref{fig:redistribute-spark-speedup-scalability2} describe the speedup and scalability of the $\bigo{(log_2N)^2}$ redistribution component in the context of platforms 1 and 2 respectively.

We note that the relative speedup of the $\bigo{(log_2N)^2}$ variant of the redistribution component (relative to the $\bigo{(log_2N)^3}$ variant) is significant in all cases: between 2 (on Platform 1) and 24 (on Platform 2). For both platforms, this speedup increases as the number of particles is increased. However, we also note that, with Platform 1 (which has a single node such that all cores share memory), the speedup decreases as the number of cores is increased for a fixed number of particles. In contrast, with platform 2, the speedup is broadly constant for large numbers of cores.

We also note that the scalability of the $\bigo{(log_2N)^2}$ variant of the redistribution component is far from ideal: increasing the number of cores culminates in minimal (if any) improvements in performance. This occurs because, in the context of both Platforms, it is the communication, and not the computation, that is limiting performance. This also explains why Platform 2's larger number of cores does not offer improved scalability relative to Platform 1: in Platform 2, the processors are distributed across multiple nodes and communicate across a network, whereas Platform 1's processors are all part of the same node and so communicate using shared memory.

\subsubsection{Resulting Overall Particle Filter Performance}

\begin{figure}
\begin{tabular}{|c|}
\hline
    \includegraphics[width=0.43\textwidth]{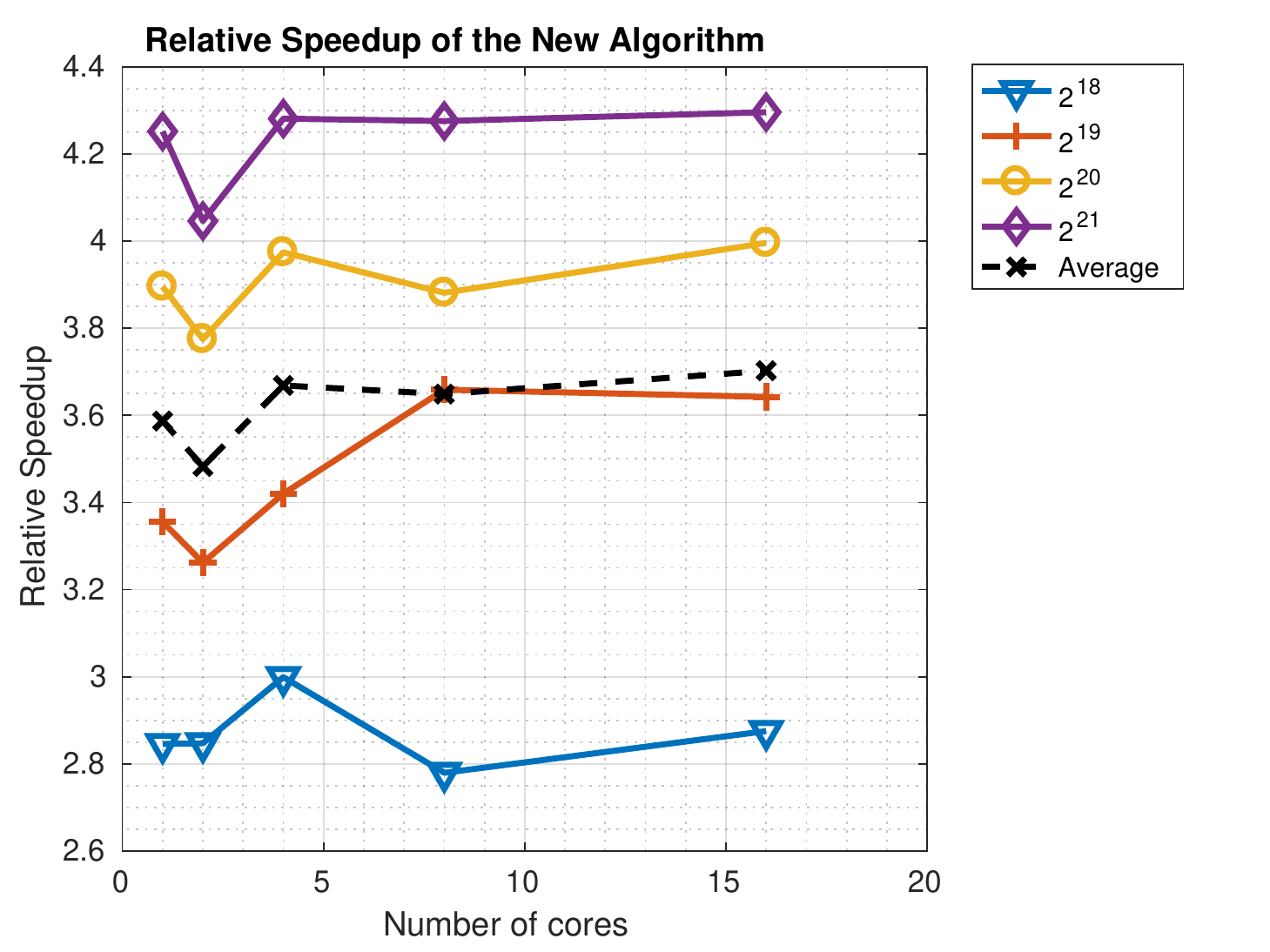}    \\
    (a) Relative Speedup of $\bigo{(log_2N)^2}$ variant\\
\hline
\end{tabular}
\quad
\begin{tabular}{|c|}
\hline
    \includegraphics[width=0.43\textwidth]{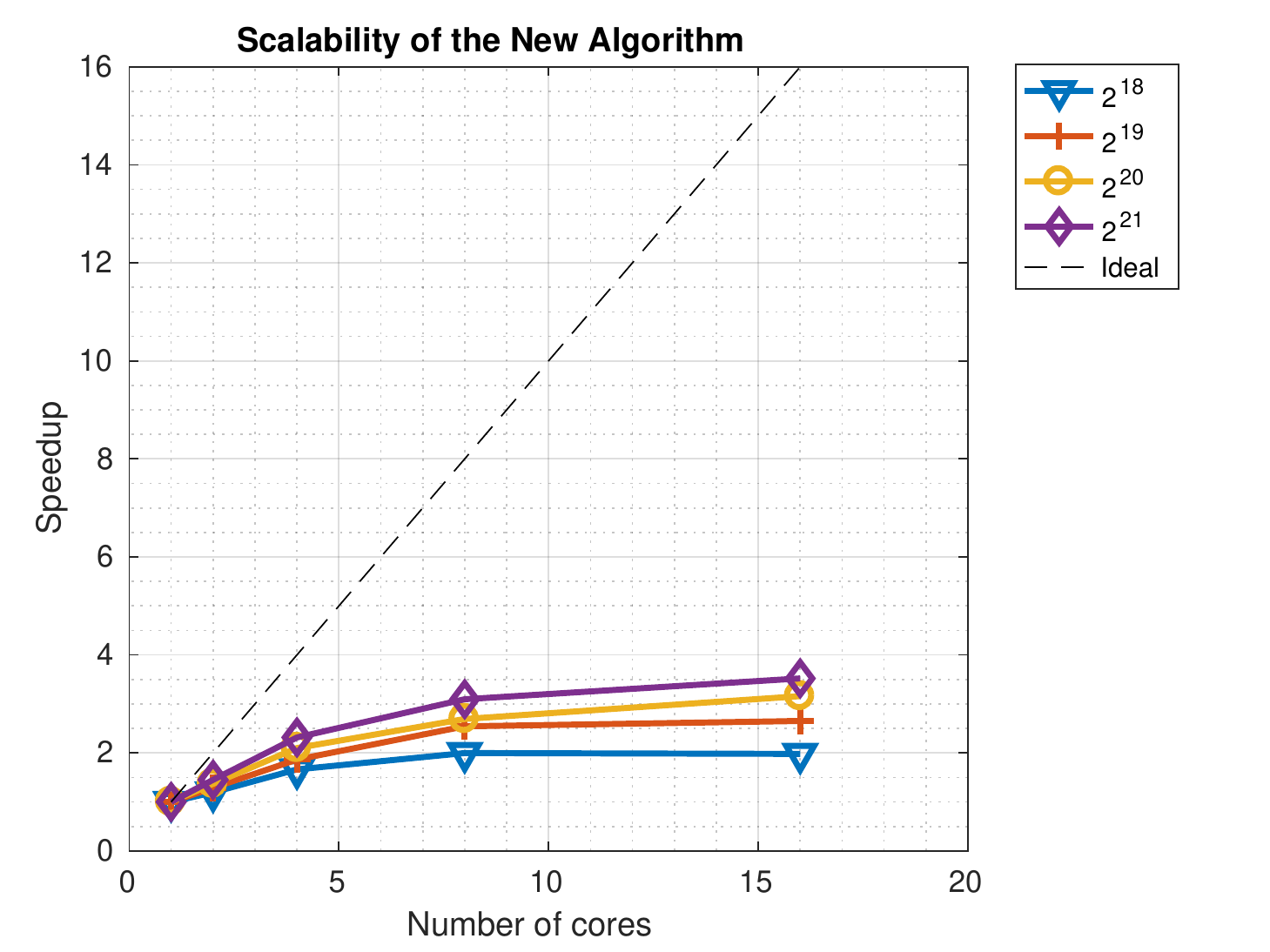} \\
    (b) Scalability of the $\bigo{(log_2N)^2}$ variant\\
\hline
\end{tabular}
\caption{Relative Speedup and Scalability of the overall particle filter algorithm using the $\bigo{(log_2N)^2}$ variant of the Redistribution component on Platform~1.}
    \label{fig:pf-spark-speedup-scalability}
\end{figure}

\begin{figure}
\begin{tabular}{|c|}
\hline
    \includegraphics[width=0.43\textwidth]{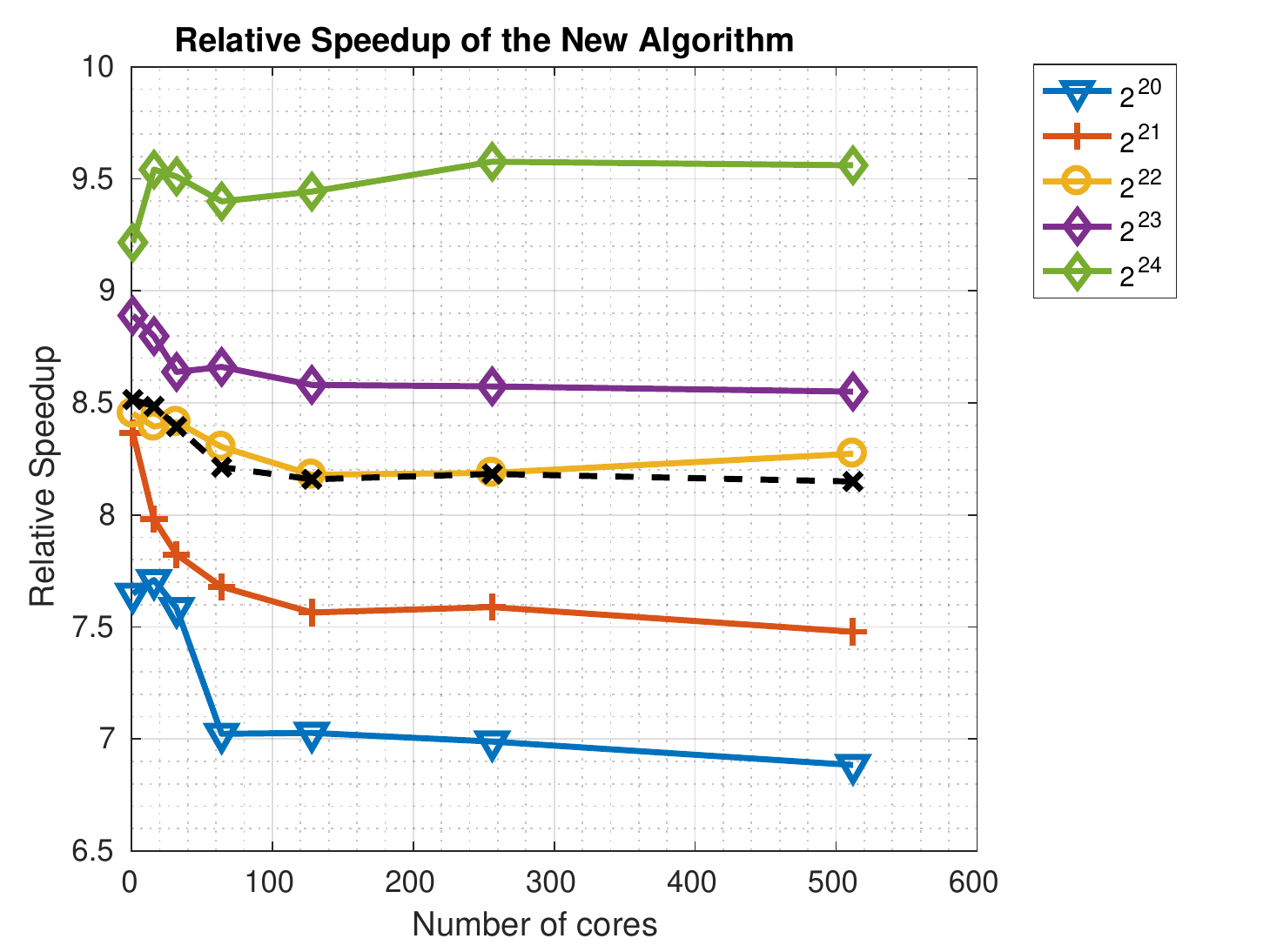} \\
    (a) Relative Speedup of $\bigo{(log_2N)^2}$ variant\\
\hline
\end{tabular}
\quad
\begin{tabular}{|c|}
\hline
    \includegraphics[width=0.43\textwidth]{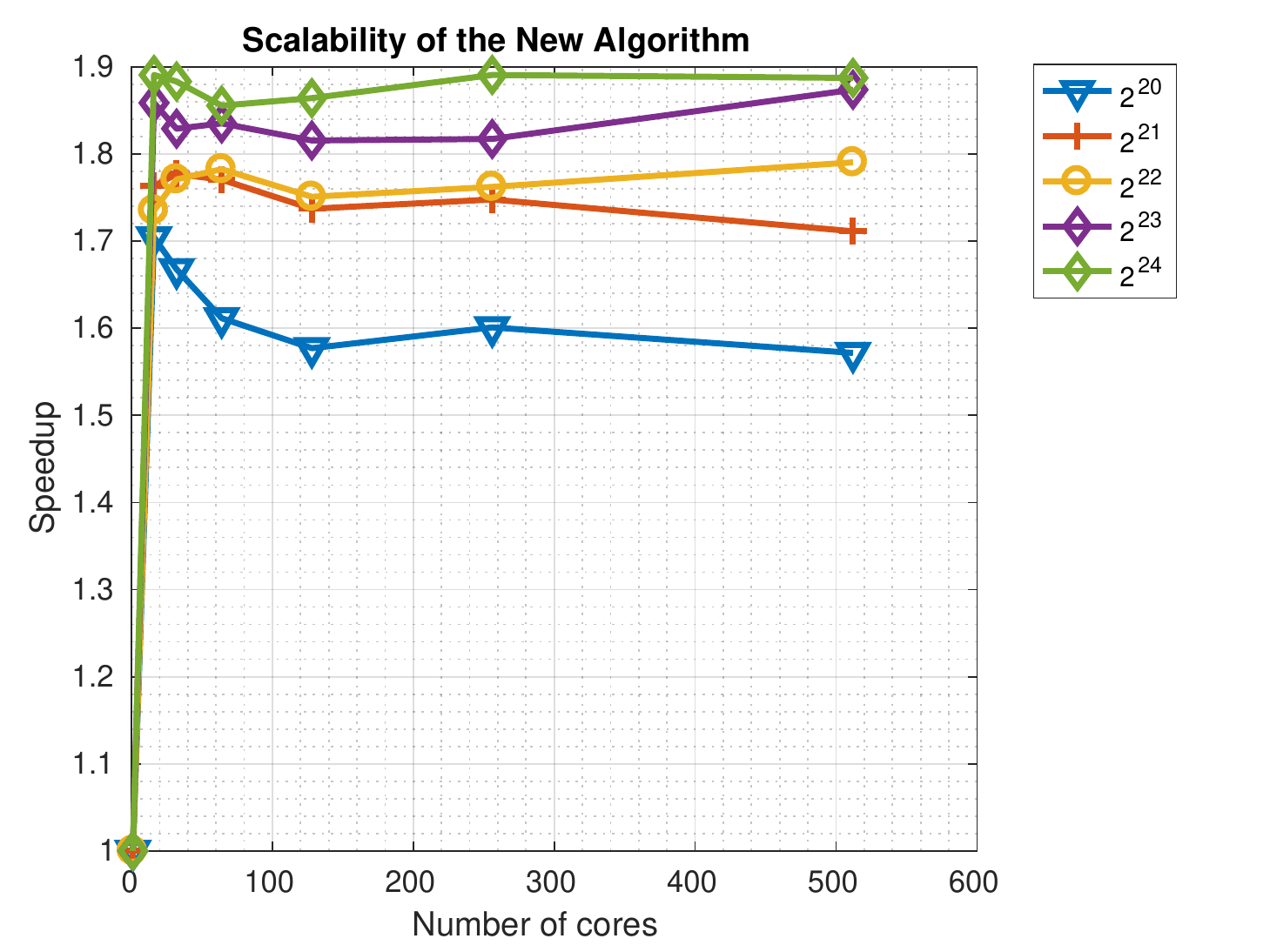} \\
    (b) Scalability of the $\bigo{(log_2N)^2}$ variant\\
\hline
\end{tabular}
\caption{Relative Speedup and Scalability of the overall particle filter algorithm using the $\bigo{(log_2N)^2}$ variant of the Redistribution component on Platform~2.}
\label{fig:pf-spark-speedup-scalability2}
\end{figure}

Figures~\ref{fig:pf-spark-speedup-scalability} and~\ref{fig:pf-spark-speedup-scalability2} describe the speedup and scalability of the overall particle filter using the $\bigo{(log_2N)^2}$ redistribution component in the context of platforms 1 and 2 respectively.

The speedups, as measured in the context of the overall particle filter algorithm are between 3 and 9.5. Again, for both platforms, the speedup increases with the number of particles. Again, the scalability is far from ideal.

\section{Related Work}
\label{sec:related}

A review of different resampling techniques is provided in~\cite{Bolic:2015}. This review makes clear that, at first sight, some of the key components of a particle filter, notably cumulative sum and redistribution, are inherently sequential\footnote{The review also highlights challenges associated with, for example, multiple processors generating independent random number sequences, discusses the relative merits of using floating-point and fixed-point numbers and points to papers discussing architecture-specific issues (e.g., in~\cite{Hong:2006,Hendeby:2010,Hwang:2013,Hendeby:2007}).}.

Indeed, this thinking has motivated research (e.g., as described in~\cite{Bolic:2005}) into approaches where a (small) number of Processing Elements (PEs) each perform local resampling and then communicate via a central process that, for example, allocates the particles to the PEs (a process that, as demonstrated in section~\ref{sec:evaluation}, results in non-deterministic run-time). In contrast to the approaches involving communication between PEs, this paper is focused on a fully distributed algorithm (with no explicit central process and so no implicit assumption of a small number of PEs).

The detailed comparison of different (single processor implementations of) resampling algorithms provided in~\cite{Hol2006} highlights that systematic resampling offers the best performance amongst the approaches considered. One strategy for parallel implementation (discussed in~\cite{Bolic:2005} and explored in more detail elsewhere\cite{murray2016parallel}) is to deliberately choose an alternative resampling algorithm such that the alternative algorithm is more amenable to parallel implementation. This paper focuses on systematic resampling specifically.

Another approach that~\cite{Bolic:2015} highlights involves each particle performing resampling using only information from its local neighbours (e.g., as described in~\cite{miguez2004new}, which, in the view of the authors, does not make obvious that, if the resampling is performed locally then the weight after resampling should be proportional to the local normalising constant\footnote{More mathematically, assume the $i$th particle has a weight (before resampling) of $w_i$ and the $j$th member of the new population is resampled as a copy of the $i$th particle with probability of $\frac{w_i}{\sum_{i'\in I_j}w_{i'}}$ where $I_j$ is the set of particles that are local to the $j$th particle. The (unnormalised) weight after resampling (based on considering the resampling process in terms of importance sampling) is $w_i\times\frac{\sum_{i'\in I_j}{w_{i'}}}{w_i}=\sum_{i'\in I_j}{w_{i'}}$. The normalised weight would then be proportional to this unnormalised weight, but scaled such that the normalised weight sums to one over all particles.}). In contrast to approaches based on considering only local neighbours, this paper describes approaches that provide exactly the same output as a single processor would have generated.

Research not explicitly covered in the aformentioned review includes the implementation described in~\cite{Maskell:SIMDPF2006} and which this paper explicitly builds upon. That implementation achieves $\bigo{(\log N)^3}$ time-complexity with $N$ parallel processors (and achieves a run-time that is not data dependent). Other related research includes (in~\cite{KirubaPF:2012}) a more complex, parallelised particle filter that uses a context-aware scheduling algorithm. They address the load imbalance arising from the na\"{i}ve parallelisation of the particle filtering by using a custom (but reusable) scheduler. In this paper, we replace the use of such a scheduler at run-time by algorithmic development at design-time.

There has been previous work on implementing particle filters in a MapReduce context (e.g., in~\cite{Bai:2013,Bai:2016}). However, this research has focused on using Hadoop and has not included a similar analysis to that documented in section~\ref{sec:evaluation} of this paper. Our analysis in that section of this paper indicates that substantial improvements are possible using Spark but also highlights that the speed-up offered using MapReduce and large numbers of processors is somewhat disappointing.

\section{Conclusions}
\label{sec:conclusions}

In this paper we have developed an improved parallel
particle filtering algorithm. The core novelty is a novel redistribution component. The component provides deterministic run-time and a time-complexity of $\bigo{(\log N)^2}$ (with $N$ particles and $N$ processors). This improves on a previous approach that achieved a time-complexity of
$\bigo{(\log N)^3}$.

A particle filter (including both the previous and new redistribution components) has been implemented using two Big Data frameworks, Hadoop and Spark. Extensive performance evaluation has been conducted. Our new component outperforms the original
version in isolation and when considering a particle filter that uses the new component in place of the original version. Our results indicate that, in the context of a particle filter, Spark's ability to perform calculations in memory enable it to offer a 25-fold improvement in run-time relative to Hadoop. Using Spark and our new component, we go on to show that, as the number of particles increases, so does the implementation efficiency.

The implementation we evaluated is limited by the communications overhead necessarily associated with giving each particle a unique key in the MapReduce framework: as a result, while we can achieve a speed-up of 3-fold with 16 cores in a single node, with 512 cores spread across 28 nodes, we only achieve a speed-up of approximately 1.4 (i.e., less). Furthermore, our implementation is outperformed by a na\"{i}ve implementation by a factor of approximately 20. Put simply, using our current implementation, we cannot yet outperform an (optimised) single processor resampling algorithm.

Of course, there will be applications where resampling is a small fraction of the total computational cost of the particle filter. In such contexts, the proposal, likelihood and/or dynamic model will be computationally demanding to calculate. These components of the particle filter are trivial to parallelise. Our future work will aim to broaden the applicability of our results beyond those applications. More specifically, we plan to focus on architectures involving a single key being related to multiple particles, explicitly minimising the need for data movement and removing the large lineages that appear to be limiting the performance possible using Spark.

Finally, we note that we have made our implementations available for
public access via an OpenSource repository at GitHub as {\tt
  particlefilter}~\cite{ParticleFilter:GitHub}.

\subsection*{Funding}
We gratefully acknowledge the second author's UK EPSRC Doctoral
Training Award.

\subsection*{Competing interests}
The authors declare that they have no competing interests.

\subsection*{Authors’ contributions}

Simon Maskell proposed the novel algorithm for redistribution in $\bigo{\left(\log N\right)^2}$ time. Jeyan Thiyagalingam led on defining the strategy for MapReduce implementation. Lykourgos Kekempanos conducted the detailed implementation and ran the simulations. All
authors contributed to the drafting of the manuscript and approved the final versions.

\section*{Acknowledgment}

We would like to acknowledge the support of STFC Daresbury and STFC Hartree
Centre for providing us with the computational resources for this work.

\bibliographystyle{spmpsci}
\bibliography{paper_all}

\end{document}